\documentclass[twocolumn]{aastex631}
\usepackage{latexsym}
\usepackage{graphicx}
\usepackage{color}
\usepackage{amsmath}

\usepackage{hyperref}
\usepackage{bm}
\usepackage{acronym}
\usepackage{url}
\usepackage{textcomp}
\usepackage{xspace}
\usepackage{acronym}
\usepackage{multirow}

\newcommand{\scinum}[2]{\ensuremath{#1\!\times\!10^{#2}}}
\newcommand{\figref}[1]{\figurename~\ref{#1}}
\newcommand{\tabref}[1]{\tablename~\ref{#1}}

\newcommand{\code}[1]{\texttt{#1}\xspace}

\begin{document}

\title{Search for continuous gravitational waves from known pulsars in the first part of the fourth LIGO-Virgo-KAGRA observing run}

%% LVK authorlist in AAS format
%\documentclass[modern]{aastex631}
%\begin{document}

%\title{LSC, Virgo and KAGRA February 2024 author list---LIGO-M2400083\\
%2024-10-11. AAS style}

\author[0000-0003-4786-2698]{A.~G.~Abac}
\affiliation{Max Planck Institute for Gravitational Physics (Albert Einstein Institute), D-14476 Potsdam, Germany}
\author{R.~Abbott}
\affiliation{LIGO Laboratory, California Institute of Technology, Pasadena, CA 91125, USA}
\author{I.~Abouelfettouh}
\affiliation{LIGO Hanford Observatory, Richland, WA 99352, USA}
\author{F.~Acernese}
\affiliation{Dipartimento di Farmacia, Universit\`a di Salerno, I-84084 Fisciano, Salerno, Italy}
\affiliation{INFN, Sezione di Napoli, I-80126 Napoli, Italy}
\author[0000-0002-8648-0767]{K.~Ackley}
\affiliation{University of Warwick, Coventry CV4 7AL, United Kingdom}
\author{S.~Adhicary}
\affiliation{The Pennsylvania State University, University Park, PA 16802, USA}
\author[0000-0002-4559-8427]{N.~Adhikari}
\affiliation{University of Wisconsin-Milwaukee, Milwaukee, WI 53201, USA}
\author[0000-0002-5731-5076]{R.~X.~Adhikari}
\affiliation{LIGO Laboratory, California Institute of Technology, Pasadena, CA 91125, USA}
\author{V.~K.~Adkins}
\affiliation{Louisiana State University, Baton Rouge, LA 70803, USA}
\author[0000-0002-8735-5554]{D.~Agarwal}
\affiliation{Universit\'e catholique de Louvain, B-1348 Louvain-la-Neuve, Belgium}
\affiliation{Inter-University Centre for Astronomy and Astrophysics, Pune 411007, India}
% ...

%\date{\today}

\begin{abstract}         
\noindent
Continuous gravitational waves (CWs) emission from neutron stars carries information about their internal structure and equation of state, and it can provide tests of General Relativity. We present a search for CWs from a set of 45 known pulsars in the first part of the fourth LIGO--Virgo--KAGRA observing run, known as O4a. We conducted a targeted search for each pulsar using three independent analysis methods considering the single-harmonic and the dual-harmonic emission models. We find no evidence of a CW signal in O4a data for both models and set upper limits on the signal amplitude and on the ellipticity, which quantifies the asymmetry in the neutron star mass distribution. 
For the single-harmonic emission model, 29 targets have the upper limit on the amplitude below the theoretical spin-down limit. The lowest upper limit on the amplitude is $6.4\!\times\!10^{-27}$ for the young energetic pulsar J0537\textminus6910, while the lowest constraint on the ellipticity is $8.8\!\times\!10^{-9}$ for the bright nearby millisecond pulsar J0437\textminus4715. Additionally, for a subset of 16 targets we performed a narrowband search that is more robust regarding the emission model, with no evidence of a signal. We also found no evidence of non-standard polarizations as predicted by the Brans-Dicke theory.
\end{abstract}

\section{Introduction}

Since their discovery in 1967, pulsars have been crucial in advancing our understanding of fundamental physics. These extremely dense and compact objects possess strong magnetic fields and rotate rapidly, emitting beams of electromagnetic (EM) radiation from hot spots near the poles or from the higher magnetosphere \citep{pulsar_rev}. 
EM observations across various wavelengths (radio, X-rays, and gamma rays) have provided detailed insights into pulsar properties, allowing precise measurements of pulsar parameters. 
Given their stability and predictability, pulsars present an excellent opportunity for the search of continuous gravitational waves (CWs) in the LIGO--Virgo--KAGRA (LVK) data.

In contrast to transient gravitational waves (GWs) emitted by binary black hole (and neutron star) mergers \citep{abbott2021gwtc2, 2021ApJ...915L...5A, 2023PhRvX..13d1039A}, CWs have yet to be observed. These signals should be nearly monochromatic, with amplitude and frequency exhibiting small variations over year-long timescales. 

CWs are expected from a time-varying non-axisymmetric mass distribution in rotating neutron stars \citep{1979PhRvD..20..351Z}.  
This could be the result of strain in the elastic crust  \citep{2000MNRAS.319..902U}, accretion from a companion star \citep{1998ApJ...501L..89B,2005ApJ...623.1044M,gittins2021modelling}, or a strong inner magnetic field \citep{1996AA...312..675B, 2002PhRvD..66h4025C}.
Alternatively, the deformation could be caused by fluid oscillations, such as those due to r-modes \citep{1998ApJ...502..708A, 1998ApJ...502..714F}.
However, the sources of CWs are likely to possess smaller mass quadrupoles compared to the sources of transient GWs, resulting in a weaker signal. Therefore, for signals to be detectable with current detectors, we need to consider nearby sources integrating long stretches (months or years worth) of detector data.

Observation of CWs from a neutron star would yield crucial insights into the star’s structure and its equation of state \citep{Haskell2023-fu, Gittins_2024}. Moreover, the form of the signal can be employed to test general relativity by measuring (or constraining) the presence of non-standard polarizations \citep{2017PhRvD..96d2001I, PhysRevD.100.104036}. Possible mechanisms of CW emission from neutron stars are discussed in greater detail by \cite{2023LRR....26....3R} and \cite{2018ASSL..457..673G}. 

The primary challenge for CW searches lies in accurately accounting for the various modulations that affect the signal received on the Earth. These include the Doppler effect due to the Earth's motion, the pulsar's rotational evolution (i.e., the slow spin-down) and relativistic effects. EM observations provide accurate measurements of the sky position and rotation parameters that allow us to predict and correct these modulations, thereby enhancing the search sensitivity. 

Based on the knowledge of the source parameters, different strategies can be used to search for CW signals in LVK data \citep[see][for a complete review of search methods]{WETTE2023102880}.
\textit{Targeted searches}, the primary subject of this paper, aim to detect CWs from known pulsars, whose timing solutions can be calculated from known rotation phases and spin-down rates. Targeted searches use full-coherent methods that integrate data over long observation times maintaining phase coherence over time. By assuming that the GW phase evolution follows the EM solution, the parameter space can be reduced to the unknown signal amplitude and polarization parameters. 
This assumption is relaxed in \textit{narrowband searches}, which are performed in a narrow band around the frequency and spin-down rate \citep[e.g.,][]{2017PhRvD..96l2006A, O2Narrowband, LIGOScientific:2021quq}. However, because these searches decrease the sensitivity and increase the computational cost, they are often performed on fewer targets.

Although several CW searches have been conducted in the last years targeting both isolated pulsars and those in binary systems, so far none of these searches have produced evidence of CWs \citep[e.g.,][]{2017ApJ...839...12A, 2019ApJ...879...10A, Abbott_2020, 2021ApJ...913L..27A, 2022ApJ...935....1A, 2021ApJ...923...85A, Nieder19, Nieder20}. In the absence of a signal, these searches set upper limits on the GW amplitude and on the ellipticity, the physical parameter that quantifies the asymmetry in the mass distribution. In many cases, these limits are more stringent than (it is said to have ``surpassed'') the so-called \textit{spin-down limit}; the theoretical limit calculated by assuming 100\% of each pulsar's spin-down luminosity to be radiated through GWs. 
In the most recent targeted search \citep{2021ApJ...913L..27A,2022ApJ...935....1A} considering O3 data, 24 pulsars surpassed their spin-down limits, including the Crab and Vela pulsars, J0537\textminus6910 and two millisecond pulsars, J0437\textminus4715 and J0711\textminus6830. Additionally, searches for an r-mode emission  are described in \citet{Rajbhandari:2021pgc} for the Crab and in \citet{2020ApJ...895...11F,Rajbhandari:2021pgc,2021arXiv210414417T} for J0537\textminus6910. Continuous improvements in detector sensitivity and data analysis techniques are progressively enhancing our ability to detect these faint signals. 

In this paper, we present a targeted search for CWs from a set of 45 known pulsars, considering LIGO data from the initial part, O4a, of the most recent LVK observing run. Pulsar selection is based on the available EM observations (see Section \ref{sec:emdata}) and on the anticipated sensitivity for targeted searches near or below the spin-down limit. Considering two different emission models, we find no evidence of CW signals in the data, and we set upper limits on the amplitude and on the ellipticity for each target. Additionally, we perform a narrowband search for a subset of 16  targets and a search for non-standard polarizations as predicted by the Brans-Dicke theory \citep{1961PhRv..124..925B}.

The paper is structured as follows. In Section \ref{sec:signal}, we briefly describe the expected signal for the emission models that we considered.
The data analysis methods used in the paper are discussed in Section \ref{sec:methods}, while in Section \ref{sec:data} we describe the EM and GW data used for the analysis. The results are summarized and discussed in detail in Sections \ref{sec:results} and \ref{sec:discussion}. Conclusions are reported in Section \ref{sec:conclusion}.

\section{Signal model}\label{sec:signal}
\subsection{Standard signal}
For the targeted search, we assume that the GW signal is locked to the rotational phase of the pulsar obtained through EM observations. For an isolated triaxial star, rotating steadily about one of its principal axes of inertia, the GW emission frequency is twice the pulsar’s spin frequency, $f_{\rm rot}$. For the single-harmonic emission model, we search for signals at $2f_{\rm rot}$. However, there are additional mechanisms which could emit GWs at other frequencies. A superfluid component beneath the crust, rotating with a spin axis misaligned to the star's rotation axis would produce an additional emission at the rotation frequency, so that overall we have a dual-harmonic emission at both once and twice the rotation frequency \citep{2010MNRAS.402.2503J}. This would not impact the EM signature. Therefore, a dual-harmonic search is performed at both $f_{\rm rot}$ and $2f_{\rm rot}$. Additionally, a single-harmonic narrowband search is performed around $2f_{\rm rot}$, allowing for the possibility of a difference in rotation rate between the pulsation-producing magnetosphere and the part of the star responsible for the CW emission.

For the general dual-harmonic emission model, the signals $h_{21}$ and $h_{22}$ at $f_{\rm rot}$ and $2f_{\rm rot}$ can be defined as \citep{2015MNRAS.453.4399P}:
\begin{eqnarray}
\nonumber
\label{eq:h21}
h_{21} &= - \frac{C_{21}}{2} \Big[&F^D_+(\alpha, \delta, \psi; t) \sin\iota \cos\iota \cos{\left(\Phi(t)+ \Phi_{21}^C\right)} +  \\
& & F^D_\times(\alpha, \delta, \psi; t)\sin\iota  \sin{\left(\Phi(t)+ \Phi_{21}^C\right)} \Big] \,,  \\
\nonumber
\label{eq:h22}
h_{22} &= - C_{22} \Big[&F^D_+(\alpha, \delta, \psi; t) (1 + \cos^2\iota) \cos{\left(2\Phi(t)+ \Phi_{22}^C\right)} +   \\
& & 2 F^D_\times(\alpha, \delta, \psi; t) \cos\iota  \sin{\left(2\Phi(t)+ \Phi_{22}^C\right)} \Big] \,,
\end{eqnarray}
where $C_{21}$ and $C_{22}$ are the dimensionless constants that give the component amplitudes, the angles $(\alpha, \delta)$ are the right ascension and declination of the source, the angles $(\iota, \psi)$ describe the orientation of the source's spin axis with respect to the observer in terms of inclination and polarization, $\Phi_{21}^C$ and $\Phi_{22}^C$ are phase angles at a defined epoch and $\Phi(t)$ is the rotational phase of the source. The antenna functions $F^D_+$ and $F^D_\times$ describe how the two polarization components (plus and cross) are projected onto the detector. These waveforms are detailed in \citet{2010MNRAS.402.2503J} and used in \citet{2022ApJ...935....1A}.

For the triaxial star described earlier, which only emits GWs at $2f_{\rm rot}$, $C_{21}$ in Equation \eqref{eq:h21} is 0, which leaves only Equation \eqref{eq:h22} which contains $C_{22}$. The amplitude $h_0$ is defined as the amplitude of the circularly polarized signal observable for a source directly above or below the plane of the detector with its spin axis pointed directly toward or away from the detector. It can be calculated as:
\begin{equation}
\label{eq:h0}
    h_0 = 2C_{22} = \frac{16\pi^2 G}{c^4} \frac{I_{zz} \varepsilon f_{\rm rot}^2 }{d} \,,
\end{equation}
where $d$ is the distance of the source. The equatorial ellipticity $\varepsilon$ for the triaxial star emitting GWs at only $2f_{\rm rot}$ is defined as 
\begin{equation}
    \varepsilon \equiv \frac{|I_{xx} - I_{yy}|}{I_{zz}} \,,
\end{equation}
where $I_{xx}$, $I_{yy}$ and $I_{zz}$ are the source's principle moments of inertia, with the star rotating about the $z$-axis. From the ellipticity, the pulsar's mass quadrupole, $Q_{22}$ can be calculated using \citep{2005PhRvL..95u1101O}
\begin{equation}
    Q_{22} = I_{zz} \varepsilon \sqrt{\frac{15}{8\pi}} .
    \label{eq:q22}
\end{equation}

The spin-down limit $h_0^{\rm sd}$ of a source is given by:
\begin{equation}
\label{eq:spin-down}
    h_0^{\rm sd}=\frac{1}{d}\left(\frac{5GI_{zz}}{2c^3}\frac{|\dot{f}_{\rm rot}|}{f_{\rm rot}}\right)^{1/2} \,
\end{equation}
where $\dot{f}_{\rm rot}$ is the rotation frequency derivative, or spin-down rate. This limit is the maximum GW amplitude allowed assuming all the lost rotational energy of the star is due to conversion into GW energy. 
It should be noted that there are two types of spin-down rates: observed, which can be affected by the transverse velocity of the source (e.g. the Shklovskii effect, described in \citet{1970SvA....13..562S}), and intrinsic. Therefore, where possible, the intrinsic spin-down rate is used in calculating the spin-down limit.

\subsection{Non-standard polarization signal}
\label{sec:sig_models_BD}

In this paper, similar to the analysis in \citet{2022ApJ...935....1A}, we search for gravitational waves (GWs) with polarizations predicted by the Brans-Dicke modification of General Relativity (GR). Brans-Dicke theory  includes two tensor polarizations, like in GR, and an additional scalar polarization. The dominant scalar radiation stems from the time-dependent dipole moment $D$. The dipole radiation occurs at the pulsar's rotational frequency $f_{\rm rot}$. Assuming the dipole moment is along the $x$-axis the amplitude $h_0^d$ of the signal is given by
\begin{equation}\label{eq:hoS}
    h_{0}^d = \frac{4 \pi G}{c^3} \zeta \frac{D f_{\rm rot}}{d}\,,
\end{equation}
where $\zeta$ is the parameter of the Brans-Dicke theory \citep[see][for details]{universe7070235}.

\section{Methods}\label{sec:methods}
In this Section, we describe the data analysis methods used in this work: three independent pipelines for the targeted searches, one pipeline for the narrowband searches and one pipeline for the non-standard polarizations searches. We use three targeted pipelines to compare independent results as our methods rely on different statistical approaches  (Bayesian or frequentist) and on different pre-processings and handling of non-stationary or non-gaussian noise disturbances in the data.

\subsection{Time-domain Bayesian Method}

The Continuous (gravitational) Wave Inference in Python (CWInPy) package is used to perform the Bayesian analysis \citep{Pitkin2022} following the method described in \cite{2019ApJ...879...10A} and summarized here. 

First, a complex, slowly evolving heterodyne is used to remove the phase evolution of the source. This includes corrections for the relative motion of the source with respect to the detector and relativistic effects \citep{Dupuis:2005}. Then, a low-pass anti-aliasing filter is applied to the data to remove the upper sideband produced from the heterodyne and limit the possibility of disturbances from spectral lines. Next, the data is down-sampled, centered about the expected signal frequency which has been shifted to 0 Hz by the heterodyne. For the dual-harmonic search, this method is repeated so that time series centered at both $f_{\rm rot}$ and $2f_{\rm rot}$ are obtained.

There are several unknown signal parameters, with the amplitude being of primary interest. Bayesian inference is used to estimate these parameters as well as the evidence for the signal model. The priors used are the same as those detailed in Appendix~2 of \citet{2017ApJ...839...12A} except for the amplitude priors, for which we use flat priors with an upper cut off much higher than the detector sensitivity. This value is \scinum{1.0}{-21} for all pulsars. The Bayesian stochastic sampling algorithm used is \textsc{dynesty} \citep{2004AIPC..735..395S, Skilling_2006}, as wrapped with \textsc{bilby} \citep{bilby_paper}, with 1024 live points (the number of points drawn from the prior).
In the absence of a signal, we calculate 95\% credible upper bounds derived from the posterior probability distributions.

\subsubsection{Restricted priors}
\label{sec:restricted}

For some pulsars, there is sufficient information to restrict our uninformative prior assumptions on their inclination and polarisation angles, for example if we have EM observations of their pulsar wind nebulae or, in the case of J1952+3252, from proper motion measurements and observations of H$\alpha$ lobes bracketing the bow shock \citep{Ng:2004}. In these cases, the parameter estimation is repeated and used along with the results from the original uninformed priors. \tabref{tab:restricted} shows the pulsars for which we could restrict the priors and the values used for the restrictions. These values were obtained using the data from Table 2 of \citet{Ng:2008} and the methods described in Appendix B of \citet{2017ApJ...839...12A}.
Each prior range is assumed to be Gaussian about the given mean and standard deviation. The two values for $\iota$ are to incorporate the unknown rotation direction in the search by using a bimodal distribution. The additional $\iota_2$ is simply $\pi -\iota_1$ radians.

\begin{table*}
    \centering
    \caption{Pulsars for which we can restrict their orientation priors using electromagnetic observations. $\Psi$ is the polarisation angle and $\iota$ is the inclination angle, with $ \iota_2 = \pi - \iota_1$.}
    \begin{tabular}{|c|c|c|c|}
        \hline
        PSR               & $\Psi$ (rad)        & $\iota_1$ (rad)     & $\iota_2$ (rad)     \\
        \hline
        J0205+6449        & 1.5760 $\pm$ 0.0078 & 1.5896 $\pm$ 0.0219 & 1.5519 $\pm$ 0.0219 \\
        J0534+2200 (Crab) & 2.1844 $\pm$ 0.0016 & 1.0850 $\pm$ 0.0149 & 2.0566 $\pm$ 0.0149 \\
        J0537\textminus6910        & 2.2864 $\pm$ 0.0383 & 1.6197 $\pm$ 0.0165 & 1.5219 $\pm$ 0.0165 \\
        J0540\textminus6919        & 2.5150 $\pm$ 0.0144 & 1.6214 $\pm$ 0.0106 & 1.5202 $\pm$ 0.0106 \\
        J0835\textminus4510 (Vela) & 2.2799 $\pm$ 0.0015 & 1.1048 $\pm$ 0.0105 & 2.0368 $\pm$ 0.0105 \\
        J1952+3252        & 0.2007 $\pm$ 0.1501 & ...                 & ...                 \\
        J2021+3651        & 0.7854 $\pm$ 0.0250 & 1.3788 $\pm$ 0.0390 & 1.7628 $\pm$ 0.0390 \\
        J2229+6114        & 1.7977 $\pm$ 0.0454 & 0.8029 $\pm$ 0.1100 & 2.3387 $\pm$ 0.1100 \\
        \hline
    \end{tabular}
    \label{tab:restricted}
\end{table*}

\subsubsection{Glitches}
\label{sec:glitches}

Glitches are transient events, where the normally stable pulsar spin suddenly increases in both rotation frequency and spin-down rate \citep{2011MNRAS.414.1679E,2013MNRAS.429..688Y,2022MNRAS.510.4049B}. These events are most common in younger, non-recycled pulsars with rarer glitches seen in some millisecond pulsars \citep{2004ApJ...612L.125C, 2016MNRAS.461.2809M}. Some searches \citep[e.g.,][]{PhysRevD.100.064058,LIGOScientific:2021quq,SGR1935} look for transient GWs in the aftermath of glitches. 

When a glitch occurs in a pulsar during the course of a GW observing run, we assume that the GW phase is affected in the same way as the EM phase. However, since the phase offset is unknown at the time of the glitch, it is incorporated into the parameter inference for the Bayesian method. This adjustment is not necessary for pulsars that glitch before or after the observation period. Only two of the pulsars in our sample experienced glitches during the length of O4a: J0537\textminus6910 with an epoch of $\sim$ MJD 60223 and J0540\textminus6919 with an epoch of $\sim$ MJD 60150 (see, e.g., \citet{2024ApJ...973L..39E, 2024ApJ...967L..13T}).

\subsection{5-vector targeted pipeline}\label{5n_targeted}
The 5-vector method is a frequentist data-analysis approach \citep{2010CQGra..27s4016A}, which has been used in the last decade in several LVK searches~\citep{2022ApJ...935....1A,Abbott_2020,2019ApJ...879...10A,2017ApJ...839...12A}. 
The 5-vector targeted pipeline relies on the Band-Sample Data (BSD)~\citep{bsd:2018} framework, i.e. a database of sub-sampled complex time-domain files (so-called BSD files) that covers $10\,$Hz and spans $1\,$month of the original dataset that allows to reduce the computational cost of the analysis.

To correct for the signal phase modulations due to, for example, the pulsar spin-down and the Doppler effect, a heterodyne method is used. Recently \citep{PhysRevD.108.122002}, the 5-vector method has been also applied to pulsars in binary systems with the implementation of the Doppler correction due to the orbital motion \citep{Leaci_2015, Leaci_2017}. After the spin-down and Doppler/relativistic corrections, there is a residual modulation at the detector due to the Earth sidereal motion that splits the frequency of the expected CW signal. 

Assuming the single-harmonic emission scenario, the central peak at twice the pulsar rotation frequency has four sidebands whose distance to the central peak is $\pm1, \pm 2 f_\oplus$ where $f_\oplus$ is the Earth's sidereal rotation frequency. A 5-vector consists of a complex array with five components corresponding to the Fourier transform of the detector's data (\textit{data 5-vector}) or of the antenna patterns (\textit{template 5-vectors} for the plus and cross component) at the five signal frequencies. 
The 5-vector method defines two matched filters in the frequency domain, defined by the normalized scalar product between the data 5-vector and the two template 5-vectors.  
To extend the analysis to $n$ detectors, the data 5-vector from each detector are combined together with coefficients that take into account the detector's sensitivity and observation time to construct the 5$n$-vectors \citep{D'Onofrio:2024Sc}. The two matched filters are linearly combined to define a detection statistic \citep{PhysRevD.108.122002}.
To assess the significance of a candidate, a p-value is computed from the noise distribution of the detection statistic using off-source frequencies as the noise background in the analyzed frequency bands. In case of no detection, a mixed Bayesian-frequentist upper limit procedure (described in \cite{2019ApJ...879...10A}) is used to set the upper limit on the amplitude, assuming a uniform prior and considering informative priors on the polarization parameters, if present (see Table \ref{tab:restricted}).  

It is not straightforward to generalize the 5-vector method to the dual-harmonic emission model due to the complexity of the expected CW signal. In this case, an additional analysis considering the emission at only the rotation frequency is performed; this would be a good approximation if the star's moment of inertia tensor is biaxial, with a small misalignment angle between its symmetry axis and the rotation axis (see \citet{2010MNRAS.402.2503J}).

\subsection{\texorpdfstring{$\mathcal{F}$}{\textit{F}}/\texorpdfstring{$\mathcal{G}$}{\textit{G}}/\texorpdfstring{$\mathcal{D}$}{\textit{D}}-statistic method}
\label{sec:fstat}
The time-domain $\mathcal{F}$/$\mathcal{G}$/$\mathcal{D}$-statistic method utilizes the $\mathcal{F}$-statistic derived in \citet{1998PhRvD..58f3001J} and the $\mathcal{G}$-statistic derived in \citet{Jaranowski:2010}. The input data for this analysis are the heterodyned data used in time-domain Bayesian analysis. The $\mathcal{F}$-statistic is employed when the amplitude, phase, and polarization of the signal are unknown, whereas the $\mathcal{G}$-statistic is applied when only the amplitude and phase are unknown, and the polarization is known (as described in Section~\ref{sec:restricted}). These methods have been used in several analyses of LIGO and Virgo data \citep{2011ApJ...737...93A, 2014ApJ...785..119A, 2017ApJ...839...12A,Abbott_2020,2022ApJ...935....1A}. Additionally, the method incorporates the $\mathcal{D}$-statistic developed in \citet{universe7070235} to search for dipole radiation in Brans-Dicke theory. The $\mathcal{D}$-statistic search was first performed in the LIGO and Virgo analysis presented in \citet{2022ApJ...935....1A}.

The three statistics are derived by calculating the maximum likelihood estimators of the signal's constant amplitude parameters. This is done by maximizing the likelihood function and then substituting the amplitude values with their respective maximum likelihood estimators. As a result, we obtain statistics that are independent of the amplitudes. In this method, a signal is detected if the value of the $\mathcal{F}$-, $\mathcal{G}$- or $\mathcal{D}$ statistic exceeds a certain threshold corresponding to an acceptable false-alarm probability. We consider a false-alarm probability of 1\% for a signal to be deemed significant. The $\mathcal{F}$-, $\mathcal{G}$-, and $\mathcal{D}$-statistics are computed for each detector and each inter-glitch period separately. The results from different detectors or inter-glitch periods are then combined incoherently by summing the respective statistics. When the values of the statistics are not statistically significant, we set an upper limit with a frequentist approach on the amplitude of the GW signal.

\subsection{5$n$-vector narrowband pipeline\label{sec:narrowband}}
The \textit{5$n$-vector narrowband pipeline} makes use of the \textit{5$n$-vector} as in \citet{narrband:2014} and follows the same principle of the method described in Section \ref{5n_targeted}.

While the former searches for a CW tightly locked to the EM emission, this assumption is here relaxed searching for CWs in a narrow frequency and spin-down range around twice the values inferred from EM observations, namely
\begin{equation}
    f\in 2f_{\rm rot} [1-\delta,\, 1+\delta]
\end{equation}
with $\delta=10^{-3}$~\citep{LIGOScientific:2021quq}, and an analogous expression for $\dot{f}$.

The method makes use of the Short Fourier Data Base (SFDB)~\citep{Astone:2005fj}, which is a collection of short-duration (2048~s) Fast Fourier Transforms (FFTs) overlapped by half.

For every target, data are Doppler corrected in the time domain with a non-uniform resampling that is independent of the CW frequency, then they are subsampled at 1~Hz.
At this point, the time series are match-filtered (using \textit{5-vectors} as for the targeted search) to estimate the two CW polarizations using a template bank in the $f-\Dot{f}$ space.
We build the template bank considering bin widths equal to the inverse of the time series duration for the frequency, while its inverse squared for the spin-down grid. Higher-order spin-down terms, if provided in the ephemerides, are fixed at twice the rotation terms to track the GW frequency evolution over time, without exploring any additional template~\citep{Astone:2014dea}.

Then, the matched filter results from different detectors are coherently combined~\citep{Mastrogiovanni:2017xjr} to evaluate the detection statistic.
From the collection of statistic values, we order them along the frequency axis to select the maximum every $10^{-4}$~Hz, marginalizing over the spin-down. We use a p-value threshold (i.e., the tail probability of the noise-only distribution) of 1\% to determine whether a selected point has to be considered an outlier. The noise-only distribution is inferred from the tail of the histogrammed non-maxima values which is fitted with an exponential~\citep{Singhal_2019}.
We use here a threshold of 1\%, similar to previous searches (e.g.,~\citet{LIGOScientific:2021quq}), after taking into account the trial factor.

If CW detection is not claimed, we set the 95\% confidence level upper limit $h_0^{95\%}$ through software-injection campaigns.

\section{Datasets}\label{sec:data}

\begin{figure}[t]
    \centering
    \includegraphics[clip,trim=0 60 0 60,scale=0.55]{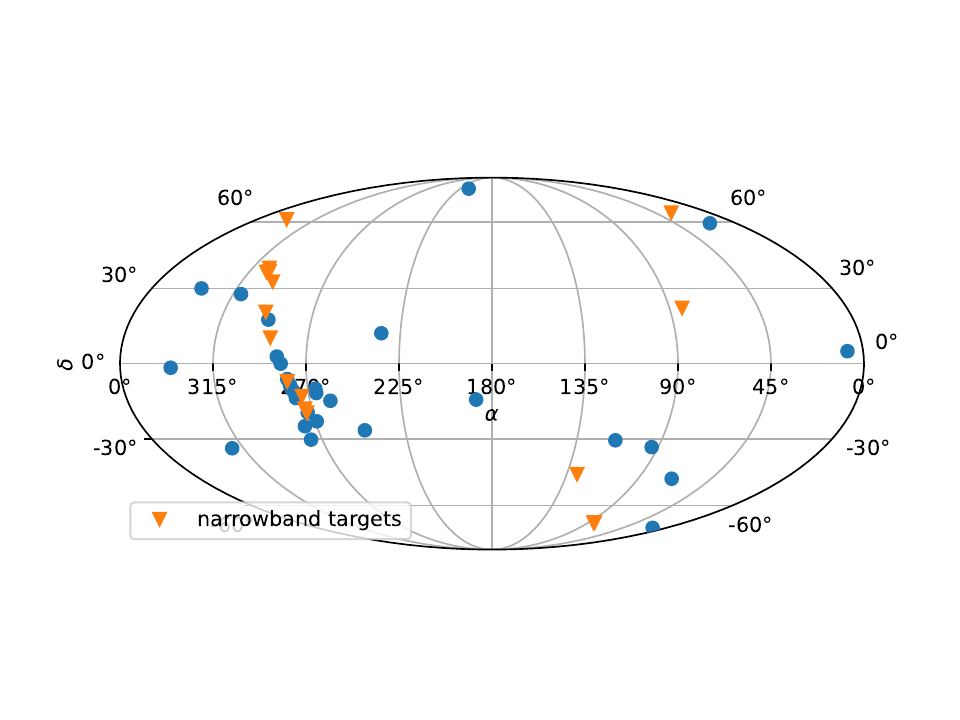}
    \caption{Sky location in equatorial coordinates of the analyzed targets. All targets are selected for the targeted search, while the triangles are the targets analyzed also with the narrowband search.}
    \label{fig:sky_pos_targets}
\end{figure}

\subsection{GW dataset}\label{sec:data_gw}

The considered dataset is the first part of the fourth observing run, known as O4a, of the LIGO Livingston (L1) and LIGO Hanford (H1) detectors\footnote{We considered \textsc{L1:GDS--CALIB\_STRAIN\_CLEAN\_AR} and \textsc{H1:GDS--CALIB\_STRAIN\_CLEAN\_AR} frame channels with \textsc{CAT1} vetoes for L1 and H1, respectively.}. 
The O4a run took place between May 24, 2023 15:00:00 UTC and ended January 16, 2024 16:00:00 UTC. The duty factors for L1 and H1 were 69.0\% and 67.5\%, respectively.  The Virgo detector has not been considered since it joined the O4 run on April 10, 2024 while the KAGRA detector is planned to join O4 by the end of the run. For a description of the upgrades to the Advanced LIGO \citep{Capote:2024rmo}, Advanced Virgo, and KAGRA detectors in preparation to the O4 run, we refer to Appendix A in \citet{Abac_2024}.

The LIGO detectors are calibrated using photon radiation pressure actuation, where an amplitude-modulated laser beam is directed onto the end test masses, causing a known change in the arm length from the equilibrium position (\citet{Karki_2016}, \citet{Viets_2018}). For the O4a data used in this analysis, the worst 1$\sigma$ calibration uncertainty is within 10\% in amplitude, and 10 degrees in phase, over the range 10–2000 Hz. The uncertainty at specific frequencies or times can be significantly smaller. 

The dataset used underwent cleaning processes depending on the data framework used by each pipeline (we refer to \citet{soni2024ligodetectorcharacterizationhalf} for more details on data quality for CW searches). 
For the Bayesian pipeline, outliers are removed before the heterodyne using a median-absolute-deviation (MAD) method as described in Chapter 3 of \cite{iglewicz1993detect} with a threshold of 3.5.
Concerning the 5-vector targeted pipeline, two cleaning steps are applied. First, short duration time-domain disturbances are identified and substracted from the data when the SFDB database - from which BSD files are built - are created (this cleaning step is shared with the 5-vector narrowband search, which also starts from the SFDB \citet{Astone:2005fj}). A second cleaning step is applied on the BSD files \citep{bsd:2018}, removing large time-domain outliers, which were not visible in the full band time series, with quadratic value larger than ten times the quadratic sum of the medians of the data real and imaginary parts (computed over non-zero samples). The $\mathcal{F}$-, $\mathcal{G}$- or $\mathcal{D}$ statistic method performs further cleaning of the fine heterodyne data through the Grubbs test (see Appendix D of \citet{2011ApJ...737...93A}).

\subsection{EM dataset}\label{sec:emdata}

The timing solutions used as inputs for the GW search were produced with EM data in the gamma rays, X-rays and radio wavelengths. The gamma ray timing solutions were obtained from Fermi-LAT \citep{2009ApJ...697.1071A}; X-ray timing solutions were obtained from Chandra \citep{2002PASP..114....1W} and the Neutron Star Interior Composition Explorer \citep[NICER,][]{lens.org/083-235-561-223-456}, while the radio timing solutions were obtained from the Nancay Radio Telescope \citep[NRT,][]{NRT2023}, the Jodrell Bank Observatory (JBO), the Argentine Institute of Radio astronomy \citep[IAR,][]{2020A&A...633A..84G}, the Mount Pleasant Radio Observatory \citep{2003ASPC..302..121L}, the Five-hundred-meter Aperture Spherical Telescope \citep[FAST,][]{2009A&A...505..919S} and the Canadian Hydrogen Intensity Mapping Experiment \citep[CHIME,][]{2021ApJS..255....5C}.

For the radio-emitting pulsars, we used \textsc{pazi} or \textsc{rfifind} tasks in PSRCHIVE \citep{2012AR&T....9..237V} and PRESTO \citep{2011ascl.soft07017R} packages respectively to mitigate Radio-frequency interferences (RFIs). Next, we folded observations with \textsc{prepfold} or \textsc{fold\_psrfits} tasks in PRESTO and PSRFITS UTILS \citep{2004PASA...21..302H} packages. 
We then cross-correlated the folded profiles with a noise-free template profile with high signal-to-noise ratio to obtain the Times of Arrivals (ToAs). Next, we select ToAs during the course of O4a run, so the solutions are valid for this time range. We use TEMPO \citep{2015ascl.soft09002N}, TEMPO2 \citep{2006MNRAS.372.1549E, 2006MNRAS.369..655H, tempo23} or PINT \citep{2019ascl.soft02007L,2021ApJ...911...45L} to characterise the rotation of each pulsar by fitting the ToAs to a Taylor series expansion:  
\begin{equation}\label{eq:timing-model}
    \phi(t)=\phi_0+ f_{\rm rot}(t-t_0)+\frac{1}{2}\dot{f}_{\rm rot}(t-t_0)^2+\frac{1}{6}\Ddot{f}_{\rm rot}(t-t_0)^3 + ...\,,
\end{equation}
\noindent where $t_0$ is the reference epoch and $\phi_0$ is the phase at $t_0$, and $f_{\rm rot}$, $\dot{f}_{\rm rot},$ and $\Ddot{f}_{\rm rot}$ are the rotation frequency of the pulsar, and its first and second derivatives, respectively. 
If higher-order derivatives are measured, we also include the corresponding terms in the Taylor expansion.

During a glitch, the rotation frequency abruptly increases. This glitch-induced alteration in the rotational phase can be taken into account in the timing model as \citep{2013MNRAS.429..688Y}:
\begin{multline}\label{eq:glitch-model}
    \phi_\mathrm{g}(t) = \Delta \phi + \Delta f_{\rm rot}^{p} (t-t_\mathrm{g}) + \frac{1}{2} \Delta \dot{f}_{\rm rot}^{\mathrm{p}} (t-t_\mathrm{g})^2 + \\ 
    \frac{1}{6} \Delta \Ddot{f}_{\rm rot}^{p}(t-t_\mathrm{g})^3 + \sum_i \left[1-\exp{\left(-\frac{t-t_\mathrm{g}}{\tau^{i}_\mathrm{d}}\right)} \right]\Delta f_{\rm rot}^{d,i}\, \tau_\mathrm{d}^{i}.
\end{multline}
The uncertainty on the glitch epoch $t_\mathrm{g}$ is counteracted here by $\Delta \phi$, and the step changes in $f_{\rm rot}$, $\dot{f}_{\rm rot}$ and  $\Delta \Ddot{f}_{\rm rot}$ at $t_\mathrm{g}$ are represented by $\Delta f_{\rm rot}^{p}$, $\Delta \dot{f}_{\rm rot}^{p}$, and $\Delta \Ddot{f}_{\rm rot}^{p}$. Finally, $\Delta f_{\rm rot}^{d,i}$ denotes temporal frequency increases that decay in $\tau_\mathrm{d}^{i}$ days.

The ToA fitting process also provides the astrometric parameters of each pulsar and the orbital parameters for binary pulsars \citep{2004hpa..book.....L}. 
Uncertainties in the values of the pulsar and orbital parameters derived from fitting the ToAs are not taken into account in targeted/narrowband searches. 

For many pulsars, their distances \citep{2004IAUS..218..139H} are based on the observed dispersion measure using the Galactic electron density distribution model YMW16 \citep{2017ApJ...835...29Y}. The uncertainties in these measurements can be as large as a factor of two. For other pulsars, the distance can be determined by measuring the parallax with the timing solution \citep{2011A&A...528A.108S} or, if the pulsar is in a binary system, the orbital period derivative \citep{2008ApJ...679..675V}. 
The first method usually results in an uncertainty ranging from 5\% to 50\% \citep{2024MNRAS.530..287S}. The second one offers significantly higher accuracy, achieving uncertainties as low as 0.1\% \citep{2008ApJ...679..675V, 2016MNRAS.455.1751R}. Other methods such as Very Long Baseline Interferometry \citep{2023ApJ...952..161L} were also used to determine pulsar distances.

For this analysis, we selected pulsars with a rotation frequency close to or greater than $10~\mathrm{Hz}$, to lie in the bandwidth of the LIGO detectors, and with an expected targeted search sensitivity for the strain amplitude within a factor 3 of the spin-down limit (see Figure \ref{fig:sky_pos_targets} for the pulsars' sky locations and Table \ref{tab:results} for the list of analyzed pulsars). 
Out of the 45 pulsars analyzed in this work, 11 pulsars belong to binary systems and there are 10 millisecond pulsars with frequencies higher than $100~\mathrm{Hz}$.

\section{Results}\label{sec:results}
\subsection{Targeted searches}
We found no statistical evidence of a CW signal in the O4a data for any of the analyzed targets. In this section, we present the results of the targeted search conducted using three different analysis methods across the full set of 45 pulsars.

The results are shown in \tabref{tab:results}. The 95\%  upper limit $h_0^{95\%}$ is given for the single-harmonic search along with the mass quadrupole $Q_{22}^{95\%}$ and ellipticity $\varepsilon^{95\%}$ upper limits calculated using the distance listed in the table and a fiducial moment of inertia $I_{zz} = 10^{38}$\,kg\,m$^2$. Uncertainties on these parameters are not taken into account, and for reference, we report the used best-fit distance values in Table \ref{tab:results} provided by the EM observation. However, for pulsars that did not surpass their spin-down limits, these $Q_{22}$ and $\epsilon$ upper limits are unphysical since they would lead to spin-down rates that are greater than their measured values. From the upper limit on the amplitude, we also compute the spin-down ratio as $h_0^{95\%}/h_0^{\text{sd}}$ with $h_0^{\text{sd}}$ defined in Equation \eqref{eq:spin-down}. The upper limits for the dual-harmonic search are included as $C_{21}^{95\%}$ and $C_{22}^{95\%}$. Finally, for the Bayesian method, the odds of a coherent signal versus incoherent noise are given for both the single $\mathcal{O}^{l=2}_{m=2}$ and dual-harmonic $\mathcal{O}^{l=2}_{m=1,2}$ searches. For the $\mathcal{F}$-statistic and for the 5-vector method, to assess the statistical significance of a candidate and quantify the consistency with the assumption of just noise, we report the p-value. 

For the two glitching pulsars, J0537\textminus6910 and J0540\textminus6919, the Bayesian results are produced when incorporating an additional phase offset in the parameter inference while for the $\mathcal{F}$-statistic and the 5-vector method, an incoherent approach is used summing the statistics from the inter-glitch periods. 
In cases with sufficient observations of the pulsar wind nebulae, results using restricted priors of inclination and polarisation angles are listed in parentheses.

\begin{figure*}
\epsscale{1.05}
\plotone{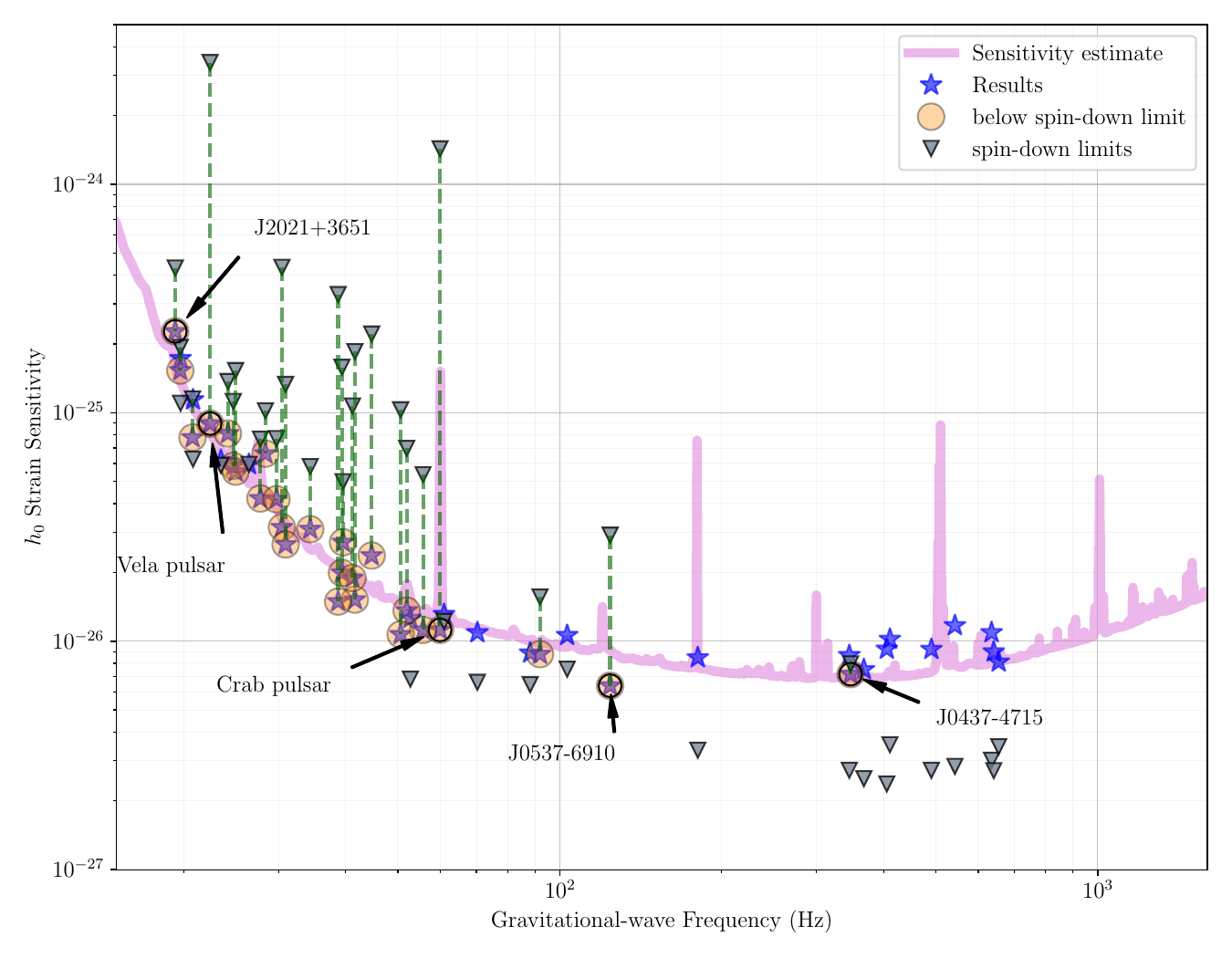}
\caption{Upper limits on $h_0$ for the 45 pulsars in this analysis using the time-domain Bayesian method and considering the single-harmonic emission model. The blue stars show 95\% credible upper limits on the amplitudes of $h_0$. Grey triangles represent the spin-down limits for each pulsar (based on the distance measurement stated in \tabref{tab:results} and assuming the canonical moment of inertia). The pink curve gives an estimate of the expected strain sensitivity of both detectors combined during the course of O4a. The upper limits from the other two pipelines are broadly consistent, as shown in Table \ref{tab:results}.}
\label{fig:h0senscurve}
\end{figure*}

\begin{figure}
\epsscale{1.15}
\plotone{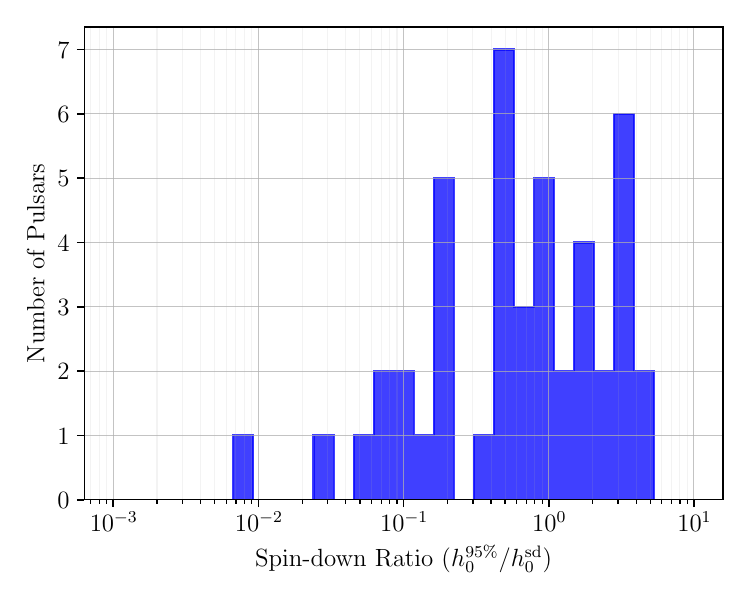}
\caption{A histogram of the spin-down ratio considering the single-harmonic emission model for 45 pulsars from the Bayesian analysis.}
\label{fig:hist}
\end{figure}

\figref{fig:h0senscurve} shows the upper limits from the Bayesian analysis for the single-harmonic search against an estimate of the sensitivity of the search using both detectors during O4a. The results for each pulsar are compared with the corresponding spin-down limit.
The results from this analysis for each pulsar are represented by the blue dots, with their corresponding spin-down limit shown by the grey triangles at the same frequency. The sensitivity curve is shown as a pink line. 
Some highlighted results for individual pulsars include the Crab pulsar (J0534+2200) which had the lowest spin-down ratio of $0.00783$, the Vela pulsar (J0835\textminus4510), J2021+3651 which had the highest odds of coherent signal versus incoherent noise with $-3.1$, J0537\textminus6910 which had the most constraining amplitude upper limit of $\scinum{6.38}{-27}$, and J0437\textminus4715 which had the most constraining ellipticity upper limit of $\scinum{8.8}{-9}$. The distribution of spin-down ratios for these results is shown in \figref{fig:hist} with 29 targets that surpass the spin-down limit and the remaining targets, which all have a spin-down ratio below 5.

\begin{figure*}
\epsscale{1.05}
\plotone{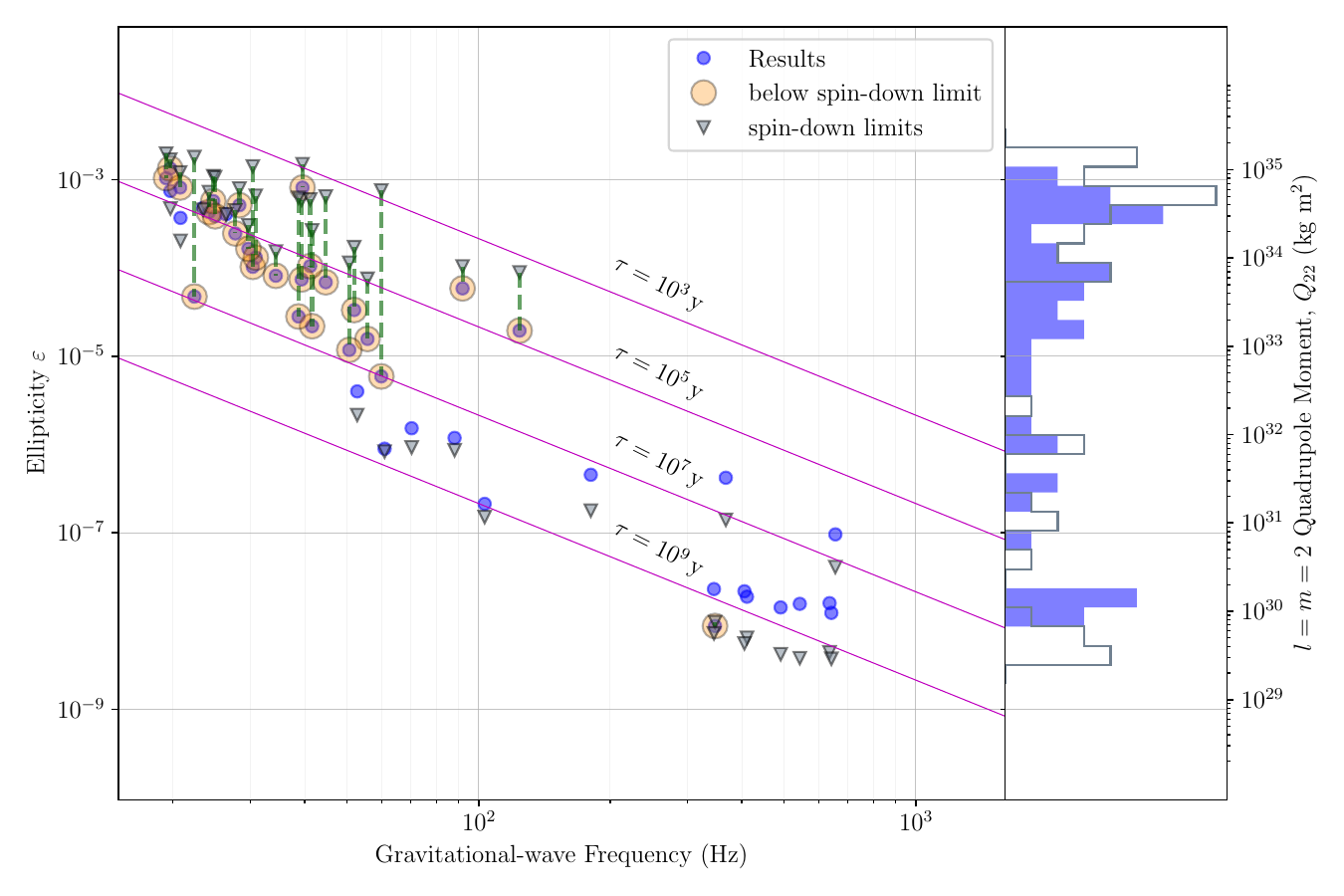}
\caption{95\% credible upper limits on ellipticity $\varepsilon^{95\%}$ and mass quadrupole $Q_{22}^{95\%}$ for all 45 pulsars using the Bayesian analysis method and considering the single-harmonic emission model. The upper limits for each pulsar are represented by blue circles while their spin-down limits are shown as grey triangles. Also included are purple contour lines of equal characteristic age $\tau = P/4\dot{P}$ assuming that GW emission alone is causing spin-down. Only the results for pulsars which surpassed their spin-down are physically meaningful. The histogram on the right shows the distribution of the ellipticities obtained from the results (blue filled bars) and in the spin-down limit (grey bars).}
\label{fig:ellfreq}
\end{figure*}

In \figref{fig:ellfreq}, the ellipticity $\varepsilon^{95\%}$ and mass quadrupole $Q_{22}^{95\%}$ upper limits are plotted against the GW frequency and compared with the corresponding spin-down limits for the ellipticity. 
The contours of equal characteristic age have been calculated using $\tau=P/4\dot{P}$ which can be derived with the assumption that GW emission alone is driving the spin-down.

As shown in Table \ref{tab:results}, there is broad agreement among the different pipelines, despite these pipelines being largely independent, and the statistical procedures used to derive the upper limits are different. The data frameworks and pre-processing procedures used by each pipeline account for the differences found in the upper limit results.

\subsection{Narrowband results}
In this section, we detail the results obtained with the narrowband pipeline, presented in Section~\ref{sec:narrowband}. The search did not highlight, for any of the 16 considered targets, outliers with a False-Alarm Probability FAP$<10^{-2}$ after taking into account the trial factor.

For the narrowband search, we consider only those targets with $h_0^{\text{sd}}$ above the expected sensitivity that is typically worse by a factor of two~\citep{narrband:2014} than that of targeted pipelines due to the trial factor. 
In this way, we selected 16 pulsars out of which 8 have not been analyzed in O3~\citep{LIGOScientific:2021quq}.
Our dataset includes the two pulsars that glitched, J0537-6910 and J0540-6919. Similar to O3, for these pulsars we split O4a data into two segments that exclude from the analysis the period around [$t_{\rm g}$-1~d, $t_{\rm g}$+2~d], with $t_{\rm g}$ the glitch epoch. The segments are then analyzed independently.

The search did not highlight any statistically significant outlier since the measured p-values are well above the threshold set by a FAP of $10^{-2}$ corrected for the trial factor. In Table~\ref{tab:narrowband_results}, we report for each target the lowest p-value found during the analysis and the threshold.

\begin{figure*}[ht]
    \centering
    \includegraphics[width=0.85\textwidth]{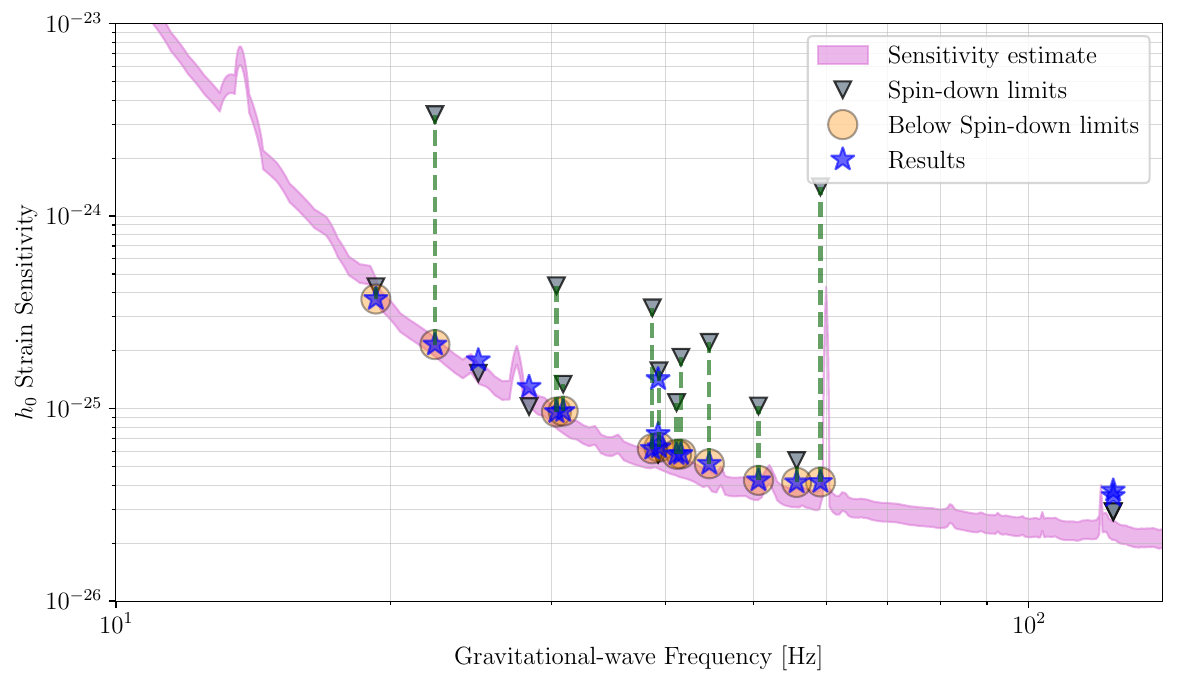}
    \caption{Expected sensitivity of the narrowband search using O4a (shaded pink region) dataset from the two LIGO detectors considering the single-harmonic emission model. The curve is compared with the spin-down limits (triangles) and the upper limits (stars) averaged over all the $10^{-4}$~Hz bands for each source. Upper limits below the spin-down limit are highlighted with orange circles.}
    \label{fig:NB_ULs}
\end{figure*}

In the absence of any detections, we have calculated upper limits at the 95\% CL for each of the analyzed targets.
Our results are listed in Table~\ref{tab:narrowband_results} and shown in Figure~\ref{fig:NB_ULs} comparing them with the expected sensitivity.

\begin{figure*}[ht]
    \centering
    \includegraphics[clip,scale=0.75]{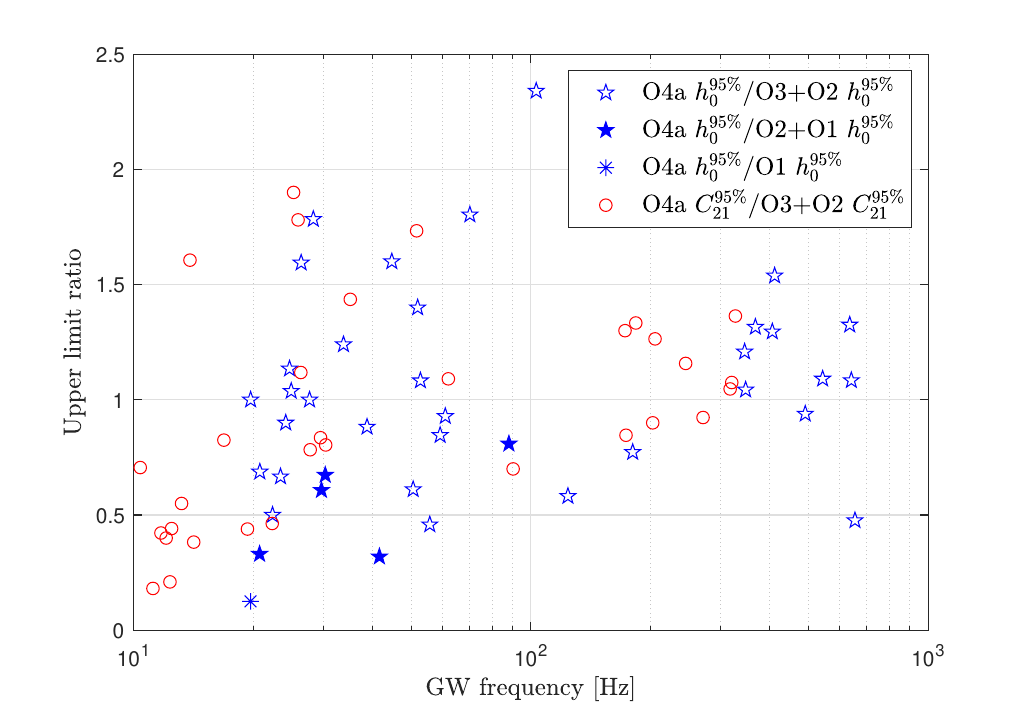}
    \caption{Blue stars show the ratio between the O4a $h_0$ upper limits for the analyzed targets (excluding the glitching pulsars) assuming the single-harmonic model divided by the corresponding $h_0$ upper limits in \citet{2022ApJ...935....1A} for the Bayesian method as a function of the corresponding frequency at twice the rotation frequency (red circles refer instead to the $C_{21}$ parameter at the rotation frequency assuming the dual-harmonic model). Blue filled stars show the $h_0$ upper limit ratios considering the targets (J0205+6449, J0737\textminus3039A, J1813\textminus1246, J1831\textminus0952, J1837\textminus0604) analyzed using O2 \citep{2019ApJ...879...10A} and O1 data (blue asterisk for J1826\textminus1334, \citet{2017ApJ...839...12A}).}
    \label{fig:UL_ratio}
\end{figure*}

\subsection{Brans-Dicke theory}

\tabref{tab:BDresults}  shows the results for the analyses on 45 pulsars using the $\mathcal{D}$-statistic to search for dipole
radiation predicted by Brans-Dicke theory. No outliers have been found in the analysis and we set upper limits on the expected amplitude defined in Equation \eqref{eq:hoS}. The upper limits in brackets shows the results using informative priors on the polarization parameters for the pulsars in Table \ref{tab:restricted}.

\section{Discussion}\label{sec:discussion}

In this section, we discuss the results in Table \ref{tab:results}. Motivated by the comparable results among the three pipelines, we consider the Bayesian pipeline as a reference. As described in the previous section, we have no evidence of a CW signal in any of the searches we conducted. 

We compare the O4a results with previous targeted searches by the LVK Collaboration \citep{2022ApJ...935....1A,Abbott_2020,2019ApJ...879...10A,2017ApJ...839...12A}, considering the first three observing runs. The ratio between the O4a upper limits on $h_0$ and $C_{21}$ and the corresponding upper limits set in previous searches is shown in Figure \ref{fig:UL_ratio}. 

34 pulsars out of the 45 considered targets in Table \ref{tab:results} have been already analyzed in the joint O2 plus O3 analysis \citep{2022ApJ...935....1A}. 
Overall the corresponding upper limits on the GW amplitude are comparable for the single-harmonic search, with some targets showing better results in O4a and some targets having worse results than those in \citet{2022ApJ...935....1A}, see Figure \ref{fig:UL_ratio}. This is expected since the targeted search sensitivity of the O2--O3 dataset is comparable to the O4a one, except at very low frequencies. The targeted search sensitivity can be expressed in terms of minimum detectable amplitude \citep{D’Onofrio_2025}, $h_{\text{min}}$. For a multi-detector analysis considering $n$ detectors and averaging over the sky position and polarization parameters,
\begin{equation}
h_{\text{min}} \approx C \sqrt{\left(\sum\limits_{i=1}^n \frac{T_i}{S_i}\right)^{-1}},
\end{equation}
where the factor $C\simeq11$ (the exact value depending on the considered pipeline), while $T_i$ and $S_i$ are respectively the effective observation time and the average power spectral density (PSD) for the $i$-th detector.  For the O4a targeted search sensitivity (with an observation time of approximately $1.3\times 10^7$ seconds for both detectors), see the pink curve in Figure \ref{fig:h0senscurve}. \clearpage
The O4 targeted searches have a sensitivity depth $\mathcal{D}=\frac{S_h/\mathrm{Hz}}{h_0}$ of around 500; here, $S_h$ is the power spectrum taking the harmonic mean of the data over time and over detectors, and $h_0$ the upper limit on the pulsar amplitude [Wette, Behnke+, Dreissigacker+].

The O4a PSDs for the two LIGO detectors are generally better, by almost a factor of $1.5$ to $2$, compared to the corresponding O3 PSD\footnote{For this estimation, we only consider the O3 run for simplicity since it dominates the combined datasets.} depending on the considered frequency band. However, the effective observation time for O4a is reduced by a factor of approximately 1.6, which diminishes the benefit of the improved detector sensitivity in O4a. As a result, we expect the sensitivity of the O4a search to be comparable to that of the combined O2+O3 search. 
Upper limits on $C_{21}$, on the other hand, are lower on average than the corresponding O2+O3 results due to a better search sensitivity at frequencies below 20 Hz. 

For the remaining targets, 5 pulsars (J0205+6449, J0737\textminus3039A, J1813\textminus1246, J1831\textminus0952, and J1837\textminus0604) have been analyzed in the O2 search \citep{2019ApJ...879...10A}, and J1826\textminus1334 has been analyzed in the O1 search \citep{2017ApJ...839...12A}. For these pulsars, we have a clear improvement in the upper limits, as shown in Figure \ref{fig:UL_ratio}. The remaining targets (J0058\textminus7218, J1811\textminus1925, J2016+3711, J2021+3651, J2022+3842) have not been analyzed in recent targeted searches and we surpass the spin-down limit for all these targets.

Many studies have been dedicated to illustrate how a future successful detection of CW will provide a wealth of information about neutron stars (see e.g.~\cite{sj_22, Lu_2023}), and even help to constrain the nuclear equation of state (see e.g.~\cite{Idrisy_2015, Ghosh_2023,Ghosh_MNRAS2023}).  It is interesting to note that with improving sensitivity of the searches with each observing run, even non-detection of a CW signal sets more and more stringent upper limits on the ellipticity and possible sources of non-axisymmetric deformations in rotating neutron stars. This may lead to a better understanding of properties of the crust, internal magnetic fields, and accretion physics \citep{1998ApJ...501L..89B,2005ApJ...623.1044M,Ciolfi:2013dta}, and even rule out certain scenarios related to r-modes or limit their maximum saturation amplitudes \citep{2021arXiv210414417T}. 

Theoretical estimates of the \emph{maximum} mountain sizes that an elastically deformed neutron star can sustain are subject to significant uncertainties, with estimates for the ellipticity $\epsilon$ ranging from  $\sim 10^{-6}$ for conventional neutron stars, to as large as $\sim 10^{-3}$ for stars with exotic solid phases (see e.g.\ \citet{2000MNRAS.319..902U, 2005PhRvL..95u1101O, hetal_07, 2013PhRvD..88d4004J, gittins2021modelling, mh_22}).  Comparison with the results given in Figure \ref{fig:ellfreq} and Table \ref{tab:results} show that our observationally-obtained  upper limits overlap with these ranges, confirming we are continuing to push into the regime of astrophysical interest.   Estimates for magnetically-induced ellipticities are similarly uncertain (see e.g.\ \citet{hsga_08, gjs_12, kotaro_etal_22}).  Nevertheless, to give a concrete example, \citet{dp_17} (see however \citet{2018MNRAS.481.4169L}) have suggested that the apparent gradual increase in the angle between the spin axis and magnetic axis of the Crab pulsar provides evidence for a magnetically-induced ellipticity $\epsilon \sim (3–10) \times 10^{-6}$.  This to be compared with our upper limit of $\epsilon \approx 6 \times 10^{-6}$ for the Crab, again confirming we are probing a regime for astrophysical interest. 

Theoretical estimates of the \emph{minimum} mountain sizes are presented in \citet{2018ApJ...863L..40W}, which provides population-based evidence for millisecond pulsars having a minimum ellipticity of $\epsilon \approx 10^{-9}$.

We stress that our upper limits are subject to the uncertainties from the detector calibration as described in Section \ref{sec:data_gw}, as well as statistical uncertainties that are dependent on the particular analysis method.  

The narrowband results in Table \ref{tab:narrowband_results} and in Figure \ref{fig:NB_ULs} show no evidence of a CW signal for the considered subset of pulsars. No outlier was found for any of the targets. Out of the 16 analyzed targets, 12 searches report an upper limit below the corresponding spin-down limit (see Table~\ref{tab:narrowband_results}), ranging from a factor of 1.16 for J2021$+$3651 to 33 for J0534$+$2200. As for other methods, the targets analyzed with O3 data~\citep{LIGOScientific:2021quq} report upper limits comparable to those in Table~\ref{tab:narrowband_results}. We stress that the narrowband search sensitivity is worse by at least a factor of 2 compared to the targeted search sensitivity as it depends on the number of templates explored for each target~\citep{Mastrogiovanni:2017xjr}.

The search for non-GR polarizations as predicted by the Brans-Dicke theory shows no evidence of a dipole radiation. The most constraining upper limit for dipole radiation is obtained for the millisecond pulsar J1719$-$1438.
Together with results from the O3 targeted search analysis \citep{2022ApJ...935....1A} the obtained upper limits constitute the first constraints on the dipole radiation from the pulsar gravitational wave observations.

\section{Conclusion}\label{sec:conclusion}
In this work, we present a search for CW signals from a set of 45 known pulsars using O4a data from the two LIGO detectors. Pulsars are chosen considering an expected sensitivity for the amplitude below or slightly above the theoretical spin-down limit  with a rotation frequency close to or greater than 10 Hz. EM observations were employed to constrain the pulsars' sky positions and rotational parameters covering the O4a data period.

We performed a targeted search utilizing three independent data analysis methods and two different emission models. No evidence of a CW signal was found for any of the targets. The upper limit results show that 29 targets surpass the theoretical spin-down limit. For 11 of the 45 pulsars not analyzed in the last LVK targeted search, we have a notable improvement in detection sensitivity compared to previous searches. For these targets, we surpass or equal the theoretical spin-down limit for the single-harmonic emission model. We also have, on average, an improvement in the upper limits for the low frequency component of the dual-harmonic search for all analyzed pulsars. For the remaining targets, the O4a upper limits are comparable to the results of the joint O2--O3 analysis described in \cite{LIGOScientific:2021quq}, which considered data with lower sensitivity but a longer observation time.

We also conducted a narrowband search for 16 pulsars and a search for non-GR polarization as predicted by Brans-Dicke theory. No evidence of a CW signal was found in any of these searches.

The analysis of the full O4 dataset will improve the sensitivity of targeted/narrowband searches for some of the pulsars analyzed in \cite{LIGOScientific:2021quq} and here, including the Crab and Vela pulsars. 
\newline \break

\section*{Acknowledgement}
% updated March 2024
%
% If not using the driver to create stand-alone pdfs for the LVC and LVK, uncomment the following lines to create the sample pdf with options spelled out
 \newif\ifcoreonly\coreonlyfalse
 \newif\ifkagra\kagratrue
 \newif\ifheader\headerfalse
\ifheader
\begin{center}{\bf\Large
\ifkagra
Conference proceedings acknowledgements for \\ the LIGO Scientific Collaboration, the Virgo Collaboration and the KAGRA Collaboration
\else
Conference proceedings acknowledgements for \\ the LIGO Scientific Collaboration and the Virgo Collaboration
\fi
}\end{center}
\fi
This material is based upon work supported by NSF's LIGO Laboratory, which is a
major facility fully funded by the National Science Foundation.
The authors also gratefully acknowledge the support of
the Science and Technology Facilities Council (STFC) of the
United Kingdom, the Max-Planck-Society (MPS), and the State of
Niedersachsen/Germany for support of the construction of Advanced LIGO 
and construction and operation of the GEO\,600 detector. 
Additional support for Advanced LIGO was provided by the Australian Research Council.
The authors gratefully acknowledge the Italian Istituto Nazionale di Fisica Nucleare (INFN),  
the French Centre National de la Recherche Scientifique (CNRS) and
the Netherlands Organization for Scientific Research (NWO)
for the construction and operation of the Virgo detector
and the creation and support  of the EGO consortium. 
The authors also gratefully acknowledge research support from these agencies as well as by 
the Council of Scientific and Industrial Research of India, 
the Department of Science and Technology, India,
the Science \& Engineering Research Board (SERB), India,
the Ministry of Human Resource Development, India,
the Spanish Agencia Estatal de Investigaci\'on (AEI),
the Spanish Ministerio de Ciencia, Innovaci\'on y Universidades,
the European Union NextGenerationEU/PRTR (PRTR-C17.I1),
the ICSC - CentroNazionale di Ricerca in High Performance Computing, Big Data
and Quantum Computing, funded by the European Union NextGenerationEU,
the Comunitat Auton\`oma de les Illes Balears through the Conselleria d'Educaci\'o i Universitats,
the Conselleria d'Innovaci\'o, Universitats, Ci\`encia i Societat Digital de la Generalitat Valenciana and
the CERCA Programme Generalitat de Catalunya, Spain,
the Polish National Agency for Academic Exchange,
the National Science Centre of Poland and the European Union - European Regional
Development Fund;
the Foundation for Polish Science (FNP),
the Polish Ministry of Science and Higher Education,
the Swiss National Science Foundation (SNSF),
the Russian Science Foundation,
the European Commission,
the European Social Funds (ESF),
the European Regional Development Funds (ERDF),
the Royal Society, 
the Scottish Funding Council, 
the Scottish Universities Physics Alliance, 
the Hungarian Scientific Research Fund (OTKA),
the French Lyon Institute of Origins (LIO),
the Belgian Fonds de la Recherche Scientifique (FRS-FNRS), 
Actions de Recherche ConcertÃ©es (ARC) and
Fonds Wetenschappelijk Onderzoek â€“ Vlaanderen (FWO), Belgium,
the Paris \^{I}le-de-France Region, 
the National Research, Development and Innovation Office of Hungary (NKFIH), 
the National Research Foundation of Korea,
the Natural Sciences and Engineering Research Council of Canada (NSERC),
the Canadian Foundation for Innovation (CFI),
the Brazilian Ministry of Science, Technology, and Innovations,
the International Center for Theoretical Physics South American Institute for Fundamental Research (ICTP-SAIFR), 
the Research Grants Council of Hong Kong,
the National Natural Science Foundation of China (NSFC),
the Israel Science Foundation (ISF),
the US-Israel Binational Science Fund (BSF),
the Leverhulme Trust, 
the Research Corporation,
the National Science and Technology Council (NSTC), Taiwan,
the United States Department of Energy,
and
the Kavli Foundation.
The authors gratefully acknowledge the support of the NSF, STFC, INFN and CNRS for provision of computational resources.

This work was supported by MEXT,
the JSPS Leading-edge Research Infrastructure Program,
JSPS Grant-in-Aid for Specially Promoted Research 26000005,
JSPS Grant-in-Aid for Scientific Research on Innovative Areas 2905: JP17H06358,
JP17H06361 and JP17H06364,
JSPS Core-to-Core Program A,
Advanced Research Networks,
JSPS Grants-in-Aid for Scientific Research (S) 17H06133 and 20H05639,
JSPS Grant-in-Aid for Transformative Research Areas (A) 20A203: JP20H05854,
the joint research program of the Institute for Cosmic Ray Research,
the University of Tokyo,
the National Research Foundation (NRF),
the Computing Infrastructure Project of Global Science experimental Data hub
Center (GSDC) at KISTI,
the Korea Astronomy and Space Science Institute (KASI),
the Ministry of Science and ICT (MSIT) in Korea,
Academia Sinica (AS),
the AS Grid Center (ASGC) and the National Science and Technology Council (NSTC)
in Taiwan under grants including the Rising Star Program and Science Vanguard
Research Program,
the Advanced Technology Center (ATC) of NAOJ,
and the Mechanical Engineering Center of KEK.

C.M.E. acknowledges support from ANID/FONDECYT, grant 1211964. S.G. acknowledges the support of the CNES. W.C.G.H. acknowledges support through grant 80NSSC22K1305 from NASA.  This work is supported by NASA through the NICER mission and the Astrophysics Explorers Program and uses data and software provided by the High Energy Astrophysics Science Archive Research Center (HEASARC), which is a service of the Astrophysics Science Division at NASA/GSFC and High Energy Astrophysics Division of the Smithsonian Astrophysical Observatory.

The Nan{\c c}ay radio Observatory is operated by the Paris
Observatory, associated with the French Centre National de la
Recherche Scientifique (CNRS), and partially supported by the Region
Centre in France. We acknowledge financial support from ``Programme
National de Cosmologie and
Galaxies'' (PNCG), and ``Programme National Hautes Energies'' (PNHE)
funded by CNRS/INSU-IN2P3-INP, CEA and CNES, France.

A.B.P. is a Banting Fellow, a McGill Space Institute~(MSI) Fellow, and a Fonds de Recherche du Quebec -- Nature et Technologies~(FRQNT) postdoctoral fellow. 

E.F. is supported by the NSF grant AST-2407399.

The activities at the Instituto Argentino de Radioastronomía (IAR) are supported by the national agency CONICET, the Province of Buenos Aires agency CIC, and the National University of La Plata (UNLP).

\software{The 5-vector method is based on the BSD framework~\citep{bsd:2018} while the narrowband method makes use of the SFDB framework~\citep{Astone:2005fj}, both of them are based on the Virgo Rome Snag software. The Bayesian analysis uses \code{CWInPy}~\citep{Pitkin2022}, which uses \code{dynesty}~\citep{2004AIPC..735..395S, Skilling_2006} within \code{bilby}~\citep{bilby_paper}. Plots are produced using \code{matplotlib}~\citep{2007CSE.....9...90H}. Many pulsar ephemerides are produced with \code{Tempo}~\citep{2015ascl.soft09002N}, \code{Tempo2}~\citep{2006MNRAS.369..655H}.}

\startlongtable
\centerwidetable
\begin{longrotatetable}
\begin{deluxetable}{lrlllllllcllcc}
\setlength\tabcolsep{2.7pt}
\tablecaption{Table of the results for the targeted search on the set of 45 known pulsars for the three considered pipelines described in Section \ref{sec:methods}. \label{tab:results}}

\tabletypesize{\scriptsize}

\tablehead{
\colhead{Pulsar Name} & 
\colhead{$f_{\text{rot}}$} & 
\colhead{$\dot{P}_{\text{rot}}$} & 
\colhead{Distance} &
\colhead{$h_0^{\text{sd}}$} &
\colhead{Analysis} & 
\colhead{$h_0^{95\%}$} &
\colhead{$\varepsilon^{95\%}$} & 
\colhead{$Q_{22}^{95\%}$} & 
\colhead{$\frac{h_0^{95\%}}{h_0^{\text{sd}}}$} & 
\colhead{$C_{21}^{95\%}$} & 
\colhead{$C_{22}^{95\%}$} & 
\colhead{Statistic\tablenotemark{@}} & 
\colhead{Statistic\tablenotemark{\#}}\\ 
\colhead{(J2000)} & 
\colhead{(Hz)} & 
\colhead{(s\,s$^{-1})$} & 
\colhead{(kpc)} & 
\colhead{~} & 
\colhead{Method} & 
\colhead{~} & 
\colhead{~} & 
\colhead{(kg\,m$^2$)} & 
\colhead{~} & 
\colhead{~} & 
\colhead{~} & 
\colhead{$^{l=2, m=1,2}$} & 
\colhead{$^{l=2, m=2}$}}

\startdata
\multirow{3}{*}{J0030+0451\tablenotemark{$\alpha$}} & \multirow{3}{*}{205.5} & \multirow{3}{*}{\scinum{-4.2}{-16}} & \multirow{3}{*}{0.3\tablenotemark{a}} &  \multirow{3}{*}{\scinum{3.5}{-27}} & Bayesian & \scinum{1.0}{-26} & \scinum{1.9}{-8} & \scinum{1.5}{30} & 2.9  & \scinum{1.1}{-26} & \scinum{5.2}{-27} & -5.1 & -10 \\  
 &  &  &  &  & $\mathcal{F}$-statistic & \scinum{1.4}{-26} & \scinum{2.7}{-8} &J0537 \scinum{2.1}{30} & 4.1 & \scinum{1.6}{-26} & \scinum{6.5}{-27} & 0.47 & 0.04 \\  
&  &  &  &  & 5$n$-vector & \scinum{9.3}{-27} & \scinum{1.7}{-8} & \scinum{1.3}{30} & 2.6 & \scinum{5.8}{-27} & \nodata & 0.61 & 0.038 \\ \hline  
\multirow{3}{*}{J0058\textminus7218\tablenotemark{$\beta$}} & \multirow{3}{*}{45.94} & \multirow{3}{*}{\scinum{-6.1}{-11}} & \multirow{3}{*}{59.70\tablenotemark{b}} & \multirow{3}{*}{\scinum{1.56}{-26}} & Bayesian &\scinum{8.8}{-27} & \scinum{5.9}{-5} & \scinum{4.5}{33} & 0.56  & \scinum{2.4}{-26} & \scinum{4.4}{-27} & -5.3 & -10 \\  
&  &  &  &  & $\mathcal{F}$-statistic & \scinum{4.9}{-27} & \scinum{3.3}{-5} & \scinum{2.5}{33} & 0.31 & \scinum{3.5}{-26} & \scinum{6.2}{-27} & 0.32 & 0.86 \\  
 &  &  &  &  & 5$n$-vector & \scinum{6.5}{-27} & \scinum{4.4}{-5} & \scinum{3.4}{33} & 0.41 & \scinum{9.4}{-27} & \nodata & 1 & 0.80 \\ \hline  
\multirow{3}{*}{J0117+5914\tablenotemark{$\gamma$}} & \multirow{3}{*}{ 9.86} & \multirow{3}{*}{\scinum{-5.69}{-13}} & \multirow{3}{*}{1.77\tablenotemark{c}} & \multirow{3}{*}{\scinum{1.1}{-25}} & Bayesian &\scinum{1.7}{-25} & \scinum{7.4}{-4} & \scinum{5.8}{34} & 1.6 &  \scinum{7.7}{-23} & \scinum{8.8}{-26} & -3.8 & -4.9 \\ 
&  &  &  &  & $\mathcal{F}$-statistic & \scinum{2.3}{-25} & \scinum{1.0}{-3} & \scinum{7.8}{34} & 2.2 & \scinum{8.4}{-23} & \scinum{1.2}{-25} & 0.39 & 0.06 \\  
 &  &  &  &  & 5$n$-vector & \scinum{3.2}{-25} & \scinum{1.4}{-3} & \scinum{1.1}{35} & 2.9 & \nodata & \nodata & \nodata & 0.87 \\ \hline  
\multirow{3}{*}{J0205+6449\tablenotemark{$\gamma$}} & \multirow{3}{*}{15.22} & \multirow{3}{*}{\scinum{-4.49}{-11}} & \multirow{3}{*}{3.20\tablenotemark{d}} & \multirow{3}{*}{\scinum{4.33}{-25}} & Bayesian & \scinum{3.2(4.2)}{-26} & \scinum{1.0(1.4)}{-4} & \scinum{8.0(10)}{33} & 0.073(0.096) &  \scinum{5.6(4.8)}{-25} & \scinum{1.5(2.0)}{-26} & -4.7(-4.5) & -8.3(-8.3) \\ 
&  &  &  &  & $\mathcal{F}$-statistic & \scinum{2.4(4.4)}{-26} & \scinum{0.8(1.5)}{-4} & \scinum{6.0(10)}{33} & 0.055(0.10) & \scinum{1.3(0.56)}{-24} & \scinum{1.1(2.2)}{-26} & 0.24(0.44) & 0.75(0.47) \\  
 &  &  &  &  & 5$n$-vector & \scinum{2.2(3.4)}{-26} & \scinum{0.72(1.1)}{-4} & \scinum{5.6(8.6)}{33} & 0.051(0.078) & \scinum{(5.7)}{-25} & \nodata & 0.86 & 0.84 \\ \hline   
\multirow{3}{*}{J0437\textminus4715\tablenotemark{$\delta$}} & \multirow{3}{*}{173.69} & \multirow{3}{*}{\scinum{-4.15}{-16}} & \multirow{3}{*}{0.16\tablenotemark{e}} & \multirow{3}{*}{\scinum{7.95}{-27}} & Bayesian & \scinum{7.2}{-27} & \scinum{8.8}{-9} & \scinum{6.8}{29} & 0.90 &  \scinum{1.1}{-26} & \scinum{3.4}{-27} & -5.4 & -10 \\ 
&  &  &  &  & $\mathcal{F}$-statistic & \scinum{9.0}{-27} & \scinum{1.1}{-8} & \scinum{8.5}{29} & 1.1 & \scinum{1.5}{-26} & \scinum{4.4}{-27} & 0.31 & 0.26 \\  
 &  &  &  &  & 5$n$-vector & \scinum{5.8}{-27} & \scinum{7.1}{-9} & \scinum{5.5}{29} & 0.32 & \scinum{5.1}{-27} & \nodata & 0.75 & 0.38 \\ \hline  
\multirow{3}{*}{J0534+2200\tablenotemark{$\gamma$}} & \multirow{3}{*}{29.95} & \multirow{3}{*}{\scinum{-3.78}{-10}} & \multirow{3}{*}{2.00\tablenotemark{f}} & \multirow{3}{*}{\scinum{1.43}{-24}} & Bayesian & \scinum{1.1(0.9)}{-26} & \scinum{5.9(5.0)}{-6} & \scinum{4.6(3.9)}{32} & 0.0078(0.0067) &  \scinum{6.6(6.0)}{-26} & \scinum{5.4(4.5)}{-27} & -5.1(-5.2) & -9.5(-9.2) \\ 
&  &  &  &  & $\mathcal{F}$-statistic & \scinum{1.5(1.2)}{-26} & \scinum{8.0(6.3)}{-6} & \scinum{6.3(4.9)}{32} & 0.011(0.0085) & \scinum{9.4(7.0)}{-26} & \scinum{8.8(6.1)}{-27} & 0.19(0.18) & 0.30(0.36) \\  
&  &  &  &  & 5$n$-vector & \scinum{1.1(1.1)}{-26} & \scinum{5.7(5.8)}{-6} & \scinum{4.3(4.4)}{32} & 0.0074(0.0075) & \scinum{2.6(2.6)}{-26} & \nodata & 0.97 & 0.44 \\ \hline  
\multirow{3}{*}{J0537\textminus6910\tablenotemark{$\beta$}} & \multirow{3}{*}{62.03} & \multirow{3}{*}{\scinum{-1.99}{-10}} & \multirow{3}{*}{49.70\tablenotemark{g}} & \multirow{3}{*}{\scinum{2.91}{-26}} & Bayesian & \scinum{6.4(8.9)}{-27} & \scinum{2.0(2.7)}{-5} & \scinum{1.5(2.1)}{33} & 0.22(0.31) & \scinum{2.4(1.4)}{-26} & \scinum{3.0(4.7)}{-27} & -5.5(-5.3) & -10.2(-10.4) \\
&  &  &  &  & $\mathcal{F}$-statistic & \scinum{8.8(4.5)}{-27} & \scinum{2.7(1.4)}{-5} & \scinum{2.0(1.0)}{33} & 0.29(0.15) & \scinum{3.2(9.0)}{-26} & \scinum{2.0(2.9)}{-27} & 0.24(0.25) & 1.00(0.92) \\  
&  &  &  &  & 5$n$-vector & \scinum{0.79(1.2)}{-26} & \scinum{2.4(3.7)}{-5} & \scinum{1.9(2.8)}{33} & 0.27(0.41) & \scinum{0.98(1.5)}{-26} & \nodata & 0.45 & 0.34 \\ \hline 
\multirow{3}{*}{J0540\textminus6919\tablenotemark{$\beta$}} & \multirow{3}{*}{19.77} & \multirow{3}{*}{\scinum{-1.87}{-10}} & \multirow{3}{*}{49.70\tablenotemark{g}} & \multirow{3}{*}{\scinum{4.99}{-26}} & Bayesian & \scinum{2.7(3.4)}{-26} & \scinum{8.1(10)}{-4} & \scinum{6.3(7.9)}{34} & 0.54(0.69) &  \scinum{1.9(1.6)}{-25} & \scinum{1.4(1.7)}{-26} & -4.6(-4.2) & -8.5(-8.4) \\ 
&  &  &  &  & $\mathcal{F}$-statistic & \scinum{2.9(2.9)}{-26} & \scinum{8.6(8.6)}{-4} & \scinum{6.7(6.7)}{34} & 0.58(0.58) & \scinum{2.9(8.8)}{-25} & \scinum{1.4(1.5)}{-26} & 0.53(0.34) & 0.32(0.45) \\ 
&  &  &  &  & 5$n$-vector & \scinum{1.6(2.4)}{-26} & \scinum{4.8(7.3)}{-4} & \scinum{3.7(5.6)}{34} & 0.27(0.41) & \scinum{3.7(5.7)}{-25} & \nodata & 0.66 & 0.76 \\ \hline 
\multirow{3}{*}{J0614\textminus3329\tablenotemark{$\alpha$}} & \multirow{3}{*}{317.59} & \multirow{3}{*}{\scinum{-1.76}{-15}} & \multirow{3}{*}{0.63\tablenotemark{h}} & \multirow{3}{*}{\scinum{3.01}{-27}} & Bayesian & \scinum{1.1}{-26} & \scinum{1.6}{-8} & \scinum{1.2}{30} & 3.6 &  \scinum{8.9}{-27} & \scinum{5.4}{-27} & -5.1 & -10 \\ 
&  &  &  &  & $\mathcal{F}$-statistic & \scinum{1.1}{-26} & \scinum{3.1}{-8} & \scinum{2.4}{30} & 3.6 & \scinum{1.0}{-26} & \scinum{5.3}{-27} & 0.66 & 0.31 \\  
 &  &  &  &  & 5$n$-vector & \scinum{8.0}{-27} & \scinum{1.2}{-8} & \scinum{9.1}{29} & 2.7 & \scinum{7.3}{-27} & \nodata & 0.15 & 0.21 \\ \hline  
\multirow{3}{*}{J0737\textminus3039A\tablenotemark{$\alpha$}} & \multirow{3}{*}{44.05} & \multirow{3}{*}{\scinum{-3.41}{-15}} & \multirow{3}{*}{1.10\tablenotemark{i}} & \multirow{3}{*}{\scinum{6.45}{-27}} & Bayesian & \scinum{8.9}{-27} & \scinum{1.2}{-6} & \scinum{9.2}{31} & 1.4 &  \scinum{1.8}{-26} & \scinum{4.3}{-27} & -5.3 & -10 \\ 
&  &  &  &  & $\mathcal{F}$-statistic & \scinum{7.5}{-27} & \scinum{1.0}{-6} & \scinum{7.6}{31} & 1.2 & \scinum{1.7}{-25} & \scinum{6.4}{-27} & 0.99 & 0.87 \\  
 &  &  &  &  & 5$n$-vector & \scinum{6.2}{-27} & \scinum{8.3}{-7} & \scinum{6.4}{31} & 0.96 & \scinum{1.4}{-26} & \nodata & 0.49 & 0.92 \\ \hline  
\multirow{3}{*}{J0835\textminus4510\tablenotemark{$\epsilon$}} & \multirow{3}{*}{11.19} & \multirow{3}{*}{\scinum{-1.57}{-11}} & \multirow{3}{*}{0.28\tablenotemark{i}} & \multirow{3}{*}{\scinum{3.41}{-24}} & Bayesian & \scinum{9.0(7.7)}{-26} & \scinum{4.7(4.1)}{-5} & \scinum{3.7(3.1)}{33} & 0.026(0.023) &  \scinum{2.9(2.4)}{-24} & \scinum{4.1(3.7)}{-26} & -4.1(-4.1) & -7.1(-7.1) \\ 
&  &  &  &  & $\mathcal{F}$-statistic & \scinum{8.2(8.2)}{-26} & \scinum{4.4(4.3)}{-5} & \scinum{3.4(3.3)}{33} & 0.024(0.024)  & \scinum{3.5(2.3)}{-24} & \scinum{4.0(3.9)}{-26} & 0.65(0.81) & 0.45(0.19) \\  
 &  &  &  &  & 5$n$-vector & \scinum{8.5(8.7)}{-26} & \scinum{4.5(4.6)}{-5} & \scinum{3.5(3.6)}{33} & 0.025(0.025) & \scinum{2.7(2.8)}{-24} & \nodata & 0.66 & 0.17 \\ \hline   
\multirow{3}{*}{J1231\textminus1411\tablenotemark{$\alpha$}} & \multirow{3}{*}{271.45} & \multirow{3}{*}{\scinum{-5.92}{-16}} & \multirow{3}{*}{0.42\tablenotemark{c}} & \multirow{3}{*}{\scinum{2.83}{-27}} & Bayesian & \scinum{1.2}{-26} & \scinum{1.6}{-8} & \scinum{1.2}{30} & 4.1 &  \scinum{1.2}{-26} & \scinum{5.9}{-27} & -4.6 & -9.9 \\ 
&  &  &  &  & $\mathcal{F}$-statistic & \scinum{1.2}{-26} & \scinum{1.6}{-8} & \scinum{1.2}{30} & 4.1 & \scinum{1.7}{-26} & \scinum{5.9}{-27} & 0.14 & 0.14 \\  
 &  &  &  &  & 5$n$-vector & \scinum{6.6}{-27} & \scinum{8.9}{-9} & \scinum{6.9}{29} & 1.4 & \scinum{5.4}{-27} & \nodata & 0.68 & 0.51 \\ \hline  
\multirow{3}{*}{J1412+7922\tablenotemark{$\beta$}} & \multirow{3}{*}{17.18} & \multirow{3}{*}{\scinum{-9.72}{-13}} & \multirow{3}{*}{3.30\tablenotemark{j}} & \multirow{3}{*}{\scinum{5.81}{-26}} & Bayesian & \scinum{3.1}{-26} & \scinum{8.2}{-5} & \scinum{6.3}{33} & 0.53  & \scinum{9.9}{-25} & \scinum{1.4}{-26} & -4.7 & -8.0 \\ 
&  &  &  &  & $\mathcal{F}$-statistic & \scinum{2.5}{-26} & \scinum{6.6}{-5} & \scinum{5.1}{33} & 0.43 & \scinum{1.2}{-24} & \scinum{1.3}{-26} & 0.17 & 0.57 \\  
&  &  &  &  & 5$n$-vector & \scinum{2.3}{-26} & \scinum{6.4}{-5} & \scinum{4.9}{33} & 0.4 & \scinum{4.0}{-25} & \nodata & 1 & 0.44 \\ \hline 
\multirow{3}{*}{J1537+1155\tablenotemark{$\alpha$}} & \multirow{3}{*}{26.38} & \multirow{3}{*}{\scinum{-1.65}{-15}} & \multirow{3}{*}{0.93\tablenotemark{k}} & \multirow{3}{*}{\scinum{6.82}{-27}} & Bayesian & \scinum{1.3}{-26} & \scinum{4.0}{-6} & \scinum{3.1}{32} & 1.8 &  \scinum{8.5}{-26} & \scinum{5.6}{-27} & -5.2 & -9.6 \\ 
&  &  &  &  & $\mathcal{F}$-statistic & \scinum{1.3}{-26} & \scinum{4.0}{-6} & \scinum{3.1}{32} & 1.8 & \scinum{7.1}{-26} & \scinum{6.4}{-27} & 0.62 & 0.91 \\  
 &  &  &  &  & 5$n$-vector & \scinum{1.6}{-26} & \scinum{5.0}{-6} & \scinum{3.9}{32} & 2.3 & \scinum{3.3}{-26} & \nodata & 0.90 & 0.22 \\ \hline  
\multirow{3}{*}{J1623\textminus2631\tablenotemark{$\gamma$}} & \multirow{3}{*}{90.29} & \multirow{3}{*}{\scinum{-5.26}{-15}} & \multirow{3}{*}{1.85\tablenotemark{l}} & \multirow{3}{*}{\scinum{3.33}{-27}} & Bayesian & \scinum{8.5}{-27} & \scinum{4.6}{-7} & \scinum{3.5}{31} & 2.6 & \scinum{1.4}{-26} & \scinum{4.1}{-27} & -5.2 & -10 \\ 
&  &  &  &  & $\mathcal{F}$-statistic & \scinum{1.0}{-26} & \scinum{5.4}{-7} & \scinum{4.1}{31} & 3.1 & \scinum{1.2}{-26} & \scinum{5.2}{-27} & 0.79 & 0.19 \\  
  &  &  &  &  & 5$n$-vector & \scinum{5.6}{-27} & \scinum{3.0}{-7} & \scinum{2.3}{31} & 1.2 & \scinum{6.8}{-27} & \nodata & 0.78 & 0.68 \\ \hline  
\multirow{3}{*}{J1719\textminus1438\tablenotemark{$\alpha$}} & \multirow{3}{*}{172.71} & \multirow{3}{*}{\scinum{-2.22}{-16}} & \multirow{3}{*}{0.34\tablenotemark{c}} & \multirow{3}{*}{\scinum{2.72}{-27}} & Bayesian & \scinum{8.7}{-27} & \scinum{2.3}{-8} & \scinum{1.8}{30} & 3.2 &  \scinum{1.3}{-26} & \scinum{3.7}{-27} & -5.3 & -10 \\ 
&  &  &  &  & $\mathcal{F}$-statistic & \scinum{7.4}{-27} & \scinum{2.0}{-8} & \scinum{1.6}{30} & 2.8 & \scinum{2.0}{-26} & \scinum{3.5}{-27} & 0.20 & 0.51 \\  
 &  &  &  &  & 5$n$-vector & \scinum{7.4}{-27} & \scinum{2.0}{-8} & \scinum{1.5}{30} & 2.6 & \scinum{7.0}{-27} & \nodata & 0.33 & 0.14 \\ \hline  
\multirow{3}{*}{J1744\textminus1134\tablenotemark{$\alpha$}} & \multirow{3}{*}{245.43} & \multirow{3}{*}{\scinum{-4.34}{-16}} & \multirow{3}{*}{0.40\tablenotemark{e}} & \multirow{3}{*}{\scinum{2.72}{-27}} & Bayesian & \scinum{9.2}{-27} & \scinum{1.4}{-8} & \scinum{1.1}{30} & 3.4 &  \scinum{1.1}{-26} & \scinum{4.4}{-27} & -5.0 & -10 \\ 
&  &  &  &  & $\mathcal{F}$-statistic & \scinum{1.4}{-26} & \scinum{2.1}{-8} & \scinum{1.7}{30} & 5.2 & \scinum{1.2}{-26} & \scinum{7.1}{-27} & 0.34 & 0.02 \\  
 &  &  &  &  & 5$n$-vector & \scinum{5.7}{-27} & \scinum{8.8}{-9} & \scinum{6.8}{29} & 1.9 & \scinum{8.8}{-27} & \nodata & 0.11 & 0.69 \\ \hline  
\multirow{3}{*}{J1745\textminus0952\tablenotemark{$\alpha$}} & \multirow{3}{*}{51.61} & \multirow{3}{*}{\scinum{-2.3}{-16}} & \multirow{3}{*}{0.23\tablenotemark{c}} & \multirow{3}{*}{\scinum{7.5}{-27}} & Bayesian  & \scinum{1.1}{-26} & \scinum{2.1}{-7} & \scinum{1.6}{31} & 1.4 &  \scinum{2.6}{-26} & \scinum{5.0}{-27} & -5.3 & -9.7 \\ 
&  &  &  &  & $\mathcal{F}$-statistic & \scinum{8.6}{-27} & \scinum{1.6}{-7} & \scinum{1.3}{31} & 1.4 & \scinum{1.6}{-26} & \scinum{4.4}{-27} & 0.66 & 0.62 \\  
 &  &  &  &  & 5$n$-vector & \scinum{1.2}{-26} & \scinum{2.5}{-7} & \scinum{1.9}{31} & 1.6 & \scinum{1.1}{-26} & \nodata & 0.91 & 0.023 \\ \hline  
\multirow{3}{*}{J1756\textminus2251\tablenotemark{$\gamma$}} & \multirow{3}{*}{35.14} & \multirow{3}{*}{\scinum{-1.26}{-15}} & \multirow{3}{*}{0.73\tablenotemark{m}} & \multirow{3}{*}{\scinum{6.6}{-27}} & Bayesian & \scinum{1.1}{-26} & \scinum{1.5}{-6} & \scinum{1.2}{32} & 1.6 &  \scinum{5.6}{-26} & \scinum{5.4}{-27} & -5.1 & -9.4 \\ 
&  &  &  &  & $\mathcal{F}$-statistic & \scinum{1.6}{-26} & \scinum{2.2}{-6} & \scinum{1.7}{32} & 2.3 & \scinum{6.3}{-26} & \scinum{7.1}{-27} & 0.17 & 0.22 \\  
 &  &  &  &  & 5$n$-vector & \scinum{1.0}{-26} & \scinum{1.5}{-6} & \scinum{1.1}{32} & 1.6 & \scinum{3.1}{-26} & \nodata & 0.031 & 0.31 \\ \hline  
\multirow{3}{*}{J1809\textminus1917\tablenotemark{$\gamma$}} & \multirow{3}{*}{12.08} & \multirow{3}{*}{\scinum{-3.73}{-12}} & \multirow{3}{*}{3.27\tablenotemark{c}} & \multirow{3}{*}{\scinum{1.37}{-25}} & Bayesian & \scinum{8.1}{-26} & \scinum{4.3}{-4} & \scinum{3.3}{34} & 0.59 &  \scinum{3.0}{-24} & \scinum{3.9}{-26} & -4.2 & -7.0 \\ 
&  &  &  &  & $\mathcal{F}$-statistic & \scinum{7.9}{-26} & \scinum{4.7}{-4} & \scinum{3.6}{34} & 0.65 & \scinum{4.3}{-24} & \scinum{3.9}{-26} & 0.52 & 0.05 \\  
 &  &  &  &  & 5$n$-vector & \scinum{4.0}{-26} & \scinum{2.1}{-4} & \scinum{1.6}{34} & 0.28 & \scinum{2.7}{-24} & \nodata & 0.079 & 0.64 \\ \hline  
\multirow{3}{*}{J1811\textminus1925\tablenotemark{$\beta$}} & \multirow{3}{*}{15.46} & \multirow{3}{*}{\scinum{-1.05}{-11}} & \multirow{3}{*}{5.00\tablenotemark{n}} & \multirow{3}{*}{\scinum{1.33}{-25}} & Bayesian & \scinum{2.6}{-26} & \scinum{1.3}{-4} & \scinum{1.0}{34} & 0.20 & \scinum{7.3}{-25} & \scinum{1.3}{-26} & -4.8 & -8.1 \\ 
&  &  &  &  & $\mathcal{F}$-statistic & \scinum{3.3}{-26}  & \scinum{1.7}{-4} & \scinum{1.3}{34} & 0.25 &  \scinum{4.3}{-25} & \scinum{1.7}{-26} & 0.67 & 0.85 \\  
  &  &  &  &  & 5$n$-vector & \scinum{3.3}{-26} & \scinum{1.6}{-4} & \scinum{1.2}{34} & 0.24 & \scinum{6.1}{-25} & \nodata & 0.23 & 0.23 \\ \hline  
\multirow{3}{*}{J1813\textminus1246\tablenotemark{$\kappa$}} & \multirow{3}{*}{20.80} & \multirow{3}{*}{\scinum{-7.6}{-12}} & \multirow{3}{*}{2.63\tablenotemark{o}} &  \multirow{3}{*}{\scinum{1.85}{-25}} & Bayesian & \scinum{1.5}{-26} & \scinum{2.2}{-5} & \scinum{1.7}{33} & 0.082 & \scinum{1.4}{-25} & \scinum{7.8}{-27} & -5.0 & -9.1 \\ 
&  &  &  &  & $\mathcal{F}$-statistic & \scinum{3.1}{-26} & \scinum{4.7}{-5} & \scinum{3.4}{33} & 0.055 & \scinum{1.3}{-25} & \scinum{1.5}{-26} & 0.72 & 0.07 \\  
 &  &  &  &  & 5$n$-vector & \scinum{1.6}{-26} & \scinum{2.3}{-5} & \scinum{1.8}{33} & 0.087 & \scinum{7.7}{-26} & \nodata & 0.73 & 0.50 \\ \hline  
\multirow{3}{*}{J1813\textminus1749\tablenotemark{$\theta$}} & \multirow{3}{*}{22.35} & \multirow{3}{*}{\scinum{-6.34}{-11}} & \multirow{3}{*}{6.15\tablenotemark{c}} & \multirow{3}{*}{\scinum{2.21}{-25}} & Bayesian & \scinum{2.4}{-26} & \scinum{6.9}{-5} & \scinum{5.3}{33} & 0.11 &  \scinum{8.8}{-26} & \scinum{1.2}{-26} & -4.4 & -8.7 \\ 
&  &  &  &  & $\mathcal{F}$-statistic & \scinum{3.1}{-26} & \scinum{8.9}{-5} & \scinum{6.8}{33} & 0.14 & \scinum{5.3}{-26} & \scinum{1.5}{-26} & 0.91 & 0.02 \\  
 &  &  &  &  & 5$n$-vector & \scinum{1.8}{-26} & \scinum{5.3}{-5} & \scinum{4.1}{33} & 0.083 & \scinum{6.6}{-26} & \nodata & 0.53 & 0.12 \\ \hline  
\multirow{3}{*}{J1823\textminus3021A\tablenotemark{$\gamma$}} & \multirow{3}{*}{183.82} & \multirow{3}{*}{\scinum{-1.14}{-13}} & \multirow{3}{*}{8.02\tablenotemark{l}} & \multirow{3}{*}{\scinum{2.5}{-27}} & Bayesian & \scinum{7.5}{-27} & \scinum{4.2}{-7} & \scinum{3.3}{31} & 3.0 &  \scinum{1.6}{-26} & \scinum{4.0}{-27} & -5.2 & -9.7 \\ 
&  &  &  &  & $\mathcal{F}$-statistic & \scinum{4.9}{-27} & \scinum{2.7}{-7} & \scinum{2.2}{31} & 2.0 & \scinum{2.3}{-26} & \scinum{3.5}{-27} & 0.12 & 0.78 \\  
 &  &  &  &  & 5$n$-vector & \scinum{4.5}{-27} & \scinum{2.5}{-7} & \scinum{1.9}{31} & 1.8 & \scinum{6.5}{-27} & \nodata & 0.40 & 0.91 \\ \hline  
 \tablebreak
\multirow{3}{*}{J1824\textminus2452A\tablenotemark{$\alpha$}} & \multirow{3}{*}{327.41} & \multirow{3}{*}{\scinum{-1.73}{-13}} & \multirow{3}{*}{5.37\tablenotemark{l}} & \multirow{3}{*}{\scinum{3.45}{-27}} & Bayesian & \scinum{8.1}{-27} & \scinum{9.6}{-8} & \scinum{7.4}{30} & 2.4 &  \scinum{1.5}{-26} & \scinum{3.7}{-27} & -5.2 & -10.0 \\ 
&  &  &  &  & $\mathcal{F}$-statistic & \scinum{7.0}{-27} & \scinum{8.3}{-8} & \scinum{6.4}{30} & 2.4 & \scinum{2.2}{-26} & \scinum{3.3}{-27} & 0.10 & 0.70 \\  
 &  &  &  &  & 5$n$-vector & \scinum{6.3}{-27} & \scinum{7.4}{-8} & \scinum{5.8}{30} & 1.8 & \scinum{7.2}{-27} & \nodata & 0.17 & 0.64 \\ \hline  
\multirow{3}{*}{J1826\textminus1334\tablenotemark{$\gamma$}} & \multirow{3}{*}{9.85} & \multirow{3}{*}{\scinum{-7.31}{-12}} & \multirow{3}{*}{3.61\tablenotemark{c}} & \multirow{3}{*}{ \scinum{1.93}{-25}} & Bayesian & \scinum{1.5}{-25} & \scinum{1.3}{-3} & \scinum{1.0}{35} & 0.79 & \scinum{4.9}{-23} & \scinum{7.0}{-26} & -3.8 & -5.7 \\ 
&  &  &  &  & $\mathcal{F}$-statistic & \scinum{1.5}{-25} & \scinum{1.3}{-3} & \scinum{1.0}{35} & 0.79 & \scinum{3.7}{-23} & \scinum{7.6}{-26} & 0.42 & 0.39 \\  
 &  &  &  &  & 5$n$-vector & \scinum{4.3}{-25} & \scinum{3.7}{-3} & \scinum{2.9}{35} & 2.2 & \nodata & \nodata & \nodata & 0.42 \\ \hline  
\multirow{3}{*}{J1828\textminus1101\tablenotemark{$\gamma$}} & \multirow{3}{*}{13.88} & \multirow{3}{*}{\scinum{-2.85}{-12}} & \multirow{3}{*}{4.77\tablenotemark{c}} & \multirow{3}{*}{\scinum{7.67}{-26}} & Bayesian & \scinum{4.2}{-26} & \scinum{2.5}{-4} & \scinum{1.9}{34} & 0.55 &  \scinum{5.3}{-24} & \scinum{2.1}{-26} & -4.6 & -7.4 \\ 
&  &  &  &  & $\mathcal{F}$-statistic & \scinum{3.5}{-26} & \scinum{2.1}{-4} & \scinum{1.6}{34} & 0.46 & \scinum{3.1}{-24} & \scinum{2.1}{-26} & 0.98 & 0.75 \\  
 &  &  &  &  & 5$n$-vector & \scinum{5.9}{-26} & \scinum{3.5}{-4} & \scinum{2.7}{34} & 0.78 & \scinum{3.8}{-24} & \nodata & 0.66 & 0.19 \\ \hline  
\multirow{3}{*}{J1831\textminus0952\tablenotemark{$\gamma$}} & \multirow{3}{*}{14.87} & \multirow{3}{*}{\scinum{-1.84}{-12}} & \multirow{3}{*}{3.68\tablenotemark{c}} & \multirow{3}{*}{\scinum{7.7}{-26}} & Bayesian & \scinum{4.2}{-26} & \scinum{1.6}{-4} & \scinum{1.3}{34} & 0.54 &  \scinum{6.4}{-25} & \scinum{1.9}{-26} & -4.6 & -8.0 \\ 
&  &  &  &  & $\mathcal{F}$-statistic & \scinum{3.2}{-26} & \scinum{1.2}{-4} & \scinum{0.99}{34} & 0.41 & \scinum{9.1}{-25} & \scinum{2.1}{-26} & 0.44 & 0.64 \\  
 &  &  &  &  & 5$n$-vector & \scinum{2.9}{-26} & \scinum{1.2}{-4} & \scinum{9.0}{33} & 0.38 & \scinum{5.2}{-25} & \nodata & 0.85 & 0.77 \\ \hline  
\multirow{3}{*}{J1833\textminus0827\tablenotemark{$\gamma$}} & \multirow{3}{*}{11.72} & \multirow{3}{*}{\scinum{-1.26}{-12}} & \multirow{3}{*}{4.50\tablenotemark{i}} & \multirow{3}{*}{\scinum{5.88}{-26}} & Bayesian & \scinum{6.2}{-26} & \scinum{4.8}{-4} & \scinum{3.7}{34} & 1.0 &  \scinum{2.7}{-24} & \scinum{3.0}{-26} & -4.4 & -7.2 \\ 
&  &  &  &  & $\mathcal{F}$-statistic & \scinum{5.4}{-26} & \scinum{4.2}{-4} & \scinum{3.2}{34} & 0.87 & \scinum{3.5}{-24} & \scinum{2.8}{-26} & 0.40 & 0.66 \\  
 &  &  &  &  & 5$n$-vector & \scinum{6.4}{-26} & \scinum{4.9}{-4} & \scinum{3.8}{34} & 1.1 & \scinum{2.4}{-24} & \nodata & 0.43 & 0.35 \\ \hline  
\multirow{3}{*}{J1837\textminus0604\tablenotemark{$\gamma$}} & \multirow{3}{*}{10.38} & \multirow{3}{*}{\scinum{-4.84}{-12}} & \multirow{3}{*}{4.78\tablenotemark{c}} & \multirow{3}{*}{\scinum{1.15}{-25}} & Bayesian & \scinum{7.8}{-26} & \scinum{8.2}{-4} & \scinum{6.3}{34} & 0.68 &  \scinum{5.6}{-24} & \scinum{3.6}{-26} & -4.4 & -7.0 \\ 
&  &  &  &  & $\mathcal{F}$-statistic & \scinum{8.8}{-26} & \scinum{9.3}{-4} & \scinum{7.1}{34} & 0.77 & \scinum{1.2}{-24} & \scinum{3.6}{-27} & 0.97 & 0.95 \\  
  &  &  &  &  & 5$n$-vector & \scinum{9.3}{-26} & \scinum{9.8}{-4} & \scinum{7.6}{34} & 0.80 & \scinum{4.7}{-24} & \nodata & 0.98 & 0.43 \\ \hline  
\multirow{3}{*}{J1838\textminus0655\tablenotemark{$\beta$}} & \multirow{3}{*}{14.18} & \multirow{3}{*}{\scinum{-9.9}{-12}} & \multirow{3}{*}{6.60\tablenotemark{p}} & \multirow{3}{*}{\scinum{1.02}{-25}} & Bayesian & \scinum{6.6}{-26} & \scinum{5.1}{-4} & \scinum{4.0}{34} & 0.65 &  \scinum{1.3}{-24} & \scinum{3.4}{-26} & -3.8 & -6.7 \\ 
&  &  &  &  & $\mathcal{F}$-statistic & \scinum{7.1}{-26} & \scinum{5.5}{-4} & \scinum{4.3}{34} & 0.70 & \scinum{2.6}{-24} & \scinum{3.6}{-26} & 0.15 & 0.04 \\  
  &  &  &  &  & 5$n$-vector & \scinum{4.2}{-26} & \scinum{3.3}{-4} & \scinum{2.5}{34} & 0.41 & \scinum{8.3}{-25} & \nodata & 0.69 & 0.29 \\ \hline  
\multirow{3}{*}{J1849\textminus0001\tablenotemark{$\beta$}} & \multirow{3}{*}{25.96} & \multirow{3}{*}{\scinum{-9.54}{-12}} & \multirow{3}{*}{7.00\tablenotemark{q}} & \multirow{3}{*}{\scinum{6.98}{-26}} & Bayesian & \scinum{1.4}{-26} & \scinum{3.3}{-5} & \scinum{2.6}{33} & 0.19 &  \scinum{1.3}{-25} & \scinum{6.0}{-27} & -5.2 & -9.1 \\ 
&  &  &  &  & $\mathcal{F}$-statistic & \scinum{1.7}{-26} & \scinum{4.0}{-5} & \scinum{3.2}{33} & 0.23 & \scinum{1.4}{-25} & \scinum{8.1}{-27} & 0.18 & 0.47 \\  
&  &  &  &  & 5$n$-vector & \scinum{1.4}{-26} & \scinum{3.4}{-5} & \scinum{2.6}{33} & 0.2 & \scinum{6.3}{-26} & \nodata & 0.078 & 0.77 \\ \hline 
\multirow{3}{*}{J1856+0245\tablenotemark{$\gamma$}} & \multirow{3}{*}{12.36} & \multirow{3}{*}{\scinum{-9.49}{-12}} & \multirow{3}{*}{6.32\tablenotemark{c}} & \multirow{3}{*}{\scinum{1.12}{-25}} & Bayesian & \scinum{5.8}{-26} & \scinum{5.7}{-4} & \scinum{4.4}{34} & 0.52 &  \scinum{1.3}{-24} & \scinum{2.9}{-26} & -4.4 & -7.8 \\ 
&  &  &  &  & $\mathcal{F}$-statistic & \scinum{5.3}{-26} & \scinum{5.2}{-4} & \scinum{4.0}{34} & 0.48 & \scinum{1.8}{-24} & \scinum{2.5}{-26} & 0.69 & 0.66 \\  
 &  &  &  &  & 5$n$-vector & \scinum{5.2}{-26} & \scinum{5.1}{-4} & \scinum{4.0}{34} & 0.47 & \scinum{1.4}{-24} & \nodata & 0.85 & 0.40 \\ \hline  
\multirow{3}{*}{J1913+1011\tablenotemark{$\gamma$}} & \multirow{3}{*}{27.85} & \multirow{3}{*}{\scinum{-2.62}{-12}} & \multirow{3}{*}{4.61\tablenotemark{c}} & \multirow{3}{*}{\scinum{5.36}{-26}} & Bayesian & \scinum{1.1}{-26} & \scinum{1.6}{-5} & \scinum{1.2}{33} & 0.21 &  \scinum{6.5}{-26} & \scinum{5.2}{-27} & -5.2 & -9.7 \\ 
&  &  &  &  & $\mathcal{F}$-statistic & \scinum{1.2}{-26} & \scinum{1.7}{-5} & \scinum{1.3}{33} & 0.23 & \scinum{4.0}{-26} & \scinum{5.8}{-27} & 0.96 & 0.60 \\  
 &  &  &  &  & 5$n$-vector & \scinum{1.2}{-26} & \scinum{1.8}{-5} & \scinum{1.4}{33} & 0.23 & \scinum{3.6}{-26} & \nodata & 0.89 & 0.35 \\ \hline  
\multirow{3}{*}{J1925+1720\tablenotemark{$\gamma$}} & \multirow{3}{*}{13.22} & \multirow{3}{*}{\scinum{-1.83}{-12}} & \multirow{3}{*}{5.05\tablenotemark{c}} & \multirow{3}{*}{\scinum{5.94}{-26}} & Bayesian & \scinum{5.9}{-26} & \scinum{4.1}{-4} & \scinum{3.1}{34} & 1.0 &  \scinum{2.2}{-24} & \scinum{2.8}{-26} & -4.2 & -6.2 \\ 
&  &  &  &  & $\mathcal{F}$-statistic & \scinum{5.4}{-26} & \scinum{3.8}{-4} & \scinum{2.8}{34} & 0.92 & \scinum{4.2}{-24} & \scinum{2.9}{-26} & 0.03 & 0.33 \\  
 &  &  &  &  & 5$n$-vector & \scinum{6.8}{-26} & \scinum{4.6}{-4} & \scinum{3.6}{34} & 1.1 & \scinum{1.5}{-24} & \nodata & 0.29 & 0.014 \\ \hline  
\multirow{3}{*}{J1935+2025\tablenotemark{$\gamma$}} & \multirow{3}{*}{12.48} & \multirow{3}{*}{\scinum{-9.47}{-12}} & \multirow{3}{*}{4.60\tablenotemark{c}} & \multirow{3}{*}{\scinum{1.53}{-25}} & Bayesian & \scinum{5.5}{-26} & \scinum{3.8}{-4} & \scinum{3.0}{34} & 0.36 &  \scinum{1.5}{-24} & \scinum{2.6}{-26} & -4.6 & -7.8 \\ 
&  &  &  &  & $\mathcal{F}$-statistic & \scinum{4.8}{-26} & \scinum{3.3}{-4} & \scinum{2.6}{34} & 0.31 & \scinum{7.7}{-25} & \scinum{2.3}{-26} & 0.98 & 0.69 \\  
 &  &  &  &  & 5$n$-vector & \scinum{3.7}{-26} & \scinum{2.6}{-4} & \scinum{2.0}{34} & 0.24 & \scinum{1.6}{-24} & \nodata & 0.57 & 0.81 \\ \hline  
 \tablebreak
\multirow{3}{*}{J1952+3252\tablenotemark{$\gamma$}} & \multirow{3}{*}{25.30} & \multirow{3}{*}{\scinum{-3.74}{-12}} & \multirow{3}{*}{3.00\tablenotemark{i}} & \multirow{3}{*}{\scinum{1.03}{-25}} & Bayesian & \scinum{1.1(1.0)}{-26} & \scinum{1.2(1.1)}{-5} & \scinum{9.2(8.2)}{32} & 0.10(0.09) & \scinum{1.9(68)}{-25} & \scinum{4.2(2.8)}{-27} & -5.3(-5.4) & -8.6(-8.2) \\
&  &  &  &  & $\mathcal{F}$-statistic & \scinum{1.0}{-26} & \scinum{1.1}{-5} & \scinum{8.4}{32} & 0.091 & \scinum{2.2}{-25} & \scinum{4.4}{-27} & 0.03 & 0.77 \\  
 &  &  &  &  & 5$n$-vector & \scinum{1.1(1.1)}{-26} & \scinum{1.2(1.2)}{-5} & \scinum{9.1(9.5)}{32} & 0.1(0.11) & \scinum{5.1(5.1)}{-26} & \nodata & 0.54 & 0.72 \\ \hline 
\multirow{3}{*}{J2016+3711\tablenotemark{$\zeta$}} & \multirow{3}{*}{19.68} & \multirow{3}{*}{\scinum{-2.81}{-11}} & \multirow{3}{*}{6.10\tablenotemark{r}} & \multirow{3}{*}{\scinum{1.58}{-25}} & Bayesian & \scinum{2.0}{-26} & \scinum{7.4}{-5} & \scinum{5.7}{33} & 0.13 &  \scinum{1.6}{-25} & \scinum{8.8}{-27} & -5.0 & -9.1 \\ 
&  &  &  &  & $\mathcal{F}$-statistic & \scinum{3.2}{-26} & \scinum{1.2}{-4} & \scinum{9.1}{33} & 0.21 & \scinum{1.2}{-25} & \scinum{1.7}{-26} & 0.84 & 0.07 \\  
&  &  &  &  & 5$n$-vector & \scinum{3.0}{-26} & \scinum{2.4}{-4} & \scinum{1.9}{34} & 0.41 & \scinum{3.8}{-25} & \nodata & 0.62 & 0.020 \\ \hline  
\multirow{3}{*}{J2021+3651\tablenotemark{$\gamma$}} & \multirow{3}{*}{9.64} & \multirow{3}{*}{\scinum{-8.89}{-12}} & \multirow{3}{*}{1.80\tablenotemark{s}} & \multirow{3}{*}{\scinum{4.3}{-25}} & Bayesian & \scinum{2.3(2.2)}{-25} & \scinum{1.0(1.0)}{-3} & \scinum{8.0(7.7)}{34} & 0.53(0.51) &  \scinum{6.0(5.0)}{-23} & \scinum{1.1(1.1)}{-25} & -3.1(-3.2) & -4.9(-5.2) \\ 
&  &  &  &  & $\mathcal{F}$-statistic & \scinum{2.2(2.1)}{-25} & \scinum{0.96(0.95)}{-3} & \scinum{7.7(7.4)}{34} & 0.51(0.49) & \scinum{2.0(1.3)}{-22} & \scinum{1.1(1.1)}{-25} & 0.09(0.65) & 0.08(0.13) \\  
 &  &  &  &  & 5$n$-vector & \scinum{2.8(4.0)}{-25} & \scinum{1.3(1.8)}{-3} & \scinum{1(1.4)}{35} & 0.66(0.93) & \nodata & \nodata & \nodata & 1 \\ \hline  
 \multirow{3}{*}{J2022+3842\tablenotemark{$\beta$}} & \multirow{3}{*}{20.59} & \multirow{3}{*}{\scinum{-3.65}{-11}} & \multirow{3}{*}{10.00\tablenotemark{t}} & \multirow{3}{*}{\scinum{1.07}{-25}} & Bayesian & \scinum{1.9}{-26} & \scinum{1.0}{-4} & \scinum{8.1}{33} & 0.18 &  \scinum{1.4}{-25} & \scinum{8.6}{-27} & -5.0 & -9.1 \\ 
&  &  &  &  & $\mathcal{F}$-statistic & \scinum{1.7}{-26} & \scinum{8.9}{-5} & \scinum{7.2}{33} & 0.16 & \scinum{8.0}{-26} & \scinum{9.0}{-27} & 0.90 & 0.60 \\  
 &  &  &  &  & 5$n$-vector & \scinum{1.4}{-26} & \scinum{7.5}{-5} & \scinum{5.8}{33} & 0.13 & \scinum{6.3}{-26} & \nodata & 1 & 0.77 \\ \hline  
\multirow{3}{*}{J2043+2740\tablenotemark{$\gamma$}} & \multirow{3}{*}{10.40} & \multirow{3}{*}{\scinum{-1.37}{-13}} & \multirow{3}{*}{1.48\tablenotemark{c}} & \multirow{3}{*}{\scinum{6.26}{-26}} & Bayesian & \scinum{1.1}{-25} & \scinum{3.7}{-4} & \scinum{2.8}{34} & 1.8 &  \scinum{1.2}{-23} & \scinum{5.2}{-26} & -4.1 & -6.0 \\ 
&  &  &  &  & $\mathcal{F}$-statistic & \scinum{1.0}{-25} & \scinum{3.4}{-4} & \scinum{2.5}{34} & 1.6 & \scinum{1.3}{-23} & \scinum{4.4}{-26} & 0.42 & 0.78 \\  
 &  &  &  &  & 5$n$-vector & \scinum{7.9}{-26} & \scinum{2.5}{-4} & \scinum{2.0}{34} & 1.2 & \scinum{6.9}{-24} & \nodata & 0.26 & 0.65 \\ \hline  
\multirow{3}{*}{J2124\textminus3358\tablenotemark{$\alpha$}} & \multirow{3}{*}{202.79} & \multirow{3}{*}{\scinum{-2.94}{-16}} & \multirow{3}{*}{0.41\tablenotemark{e}} & \multirow{3}{*}{\scinum{2.37}{-27}} & Bayesian & \scinum{9.2}{-27} & \scinum{2.2}{-8} & \scinum{1.7}{30} & 3.9 &  \scinum{9.9}{-27} & \scinum{4.8}{-27} & -5.0 & -10 \\ 
&  &  &  &  & $\mathcal{F}$-statistic & \scinum{9.7}{-27} & \scinum{2.3}{-8} & \scinum{1.8}{30} & 4.1 & \scinum{9.5}{-27} & \scinum{4.8}{-27} & 0.78 & 0.23 \\  
 &  &  &  &  & 5$n$-vector & \scinum{5.4}{-27} & \scinum{1.3}{-8} & \scinum{9.8}{29} & 1.3 & \scinum{5.1}{-27} & \nodata & 0.73 & 0.62 \\ \hline  
\multirow{3}{*}{J2214+3000\tablenotemark{$\eta$}} & \multirow{3}{*}{320.59} & \multirow{3}{*}{\scinum{-1.31}{-15}} & \multirow{3}{*}{0.60\tablenotemark{u}} & \multirow{3}{*}{\scinum{2.71}{-27}} & Bayesian & \scinum{9.0}{-27} & \scinum{1.2}{-8} & \scinum{9.6}{29} & 3.3 &  \scinum{8.6}{-27} & \scinum{4.6}{-27} & -5.3 & -10 \\ 
&  &  &  &  & $\mathcal{F}$-statistic & \scinum{4.5}{-27} & \scinum{0.60}{-8} & \scinum{4.8}{29} & 1.7 & \scinum{9.0}{-27} & \scinum{5.8}{-27} & 0.81 & 0.87 \\  
 &  &  &  &  & 5$n$-vector & \scinum{5.4}{-27} & \scinum{7.4}{-9} & \scinum{5.7}{29} & 1.8 & \scinum{7.5}{-27} & \nodata & 0.11 & 0.84 \\ \hline  
\multirow{3}{*}{J2222\textminus0137\tablenotemark{$\alpha$}} & \multirow{3}{*}{30.47} & \multirow{3}{*}{\scinum{-4.99}{-16}} & \multirow{3}{*}{0.27\tablenotemark{v}} & \multirow{3}{*}{\scinum{1.22}{-26}} & Bayesian & \scinum{1.3}{-26} & \scinum{9.0}{-7} & \scinum{6.9}{31} & 1.1 &  \scinum{4.1}{-26} & \scinum{6.4}{-27} & -5.1 & -9.7 \\ 
&  &  &  &  & $\mathcal{F}$-statistic & \scinum{2.2}{-26} & \scinum{1.5}{-6} & \scinum{1.2}{32} & 1.9 & \scinum{5.0}{-26} & \scinum{1.1}{-26} & 0.93 & 0.06 \\  
  &  &  &  &  & 5$n$-vector & \scinum{1.3}{-26} & \scinum{9.1}{-7} & \scinum{7.0}{31} & 3.3 & \scinum{4.9}{-26} & \nodata & 0.016 & 0.15 \\ \hline  
\multirow{3}{*}{J2229+6114\tablenotemark{$\gamma$}} & \multirow{3}{*}{19.36} & \multirow{3}{*}{\scinum{-2.9}{-11}} & \multirow{3}{*}{3.00\tablenotemark{w}} & \multirow{3}{*}{\scinum{3.29}{-25}} & Bayesian & \scinum{1.5(0.9)}{-26} & \scinum{2.8(1.8)}{-5} & \scinum{2.2(1.4)}{33} & 0.045(0.028) &  \scinum{1.8(1.6)}{-25} & \scinum{6.8(4.6)}{-27} & -5.1(-5.3) & -9.2(-9.4) \\
&  &  &  &  & $\mathcal{F}$-statistic & \scinum{1.5(0.63)}{-26} & \scinum{2.9(1.2)}{-5} & \scinum{2.2(0.98)}{33} & 0.046(0.032) & \scinum{2.9(2.5)}{-25} & \scinum{4.7(2.5)}{-27} & 0.24(0.41) & 0.99(0.99) \\  
 &  &  &  &  & 5$n$-vector & \scinum{1.3(0.99)}{-26} & \scinum{2.4(1.9)}{-5} & \scinum{1.8(1.4)}{33} & 0.037(0.029) & \scinum{3.9(3.2)}{-27} & \nodata & 0.75 & 0.88 \\ \hline  
\enddata
\tablerefs{
The last two columns refers to the significance of the data against the noise hypothesis ($@$ for the dual-harmonic emission model, $\#$ for the single-harmonic emission model).  For the Bayesian method, the columns show the base-10 logarithm of the Bayesian odds, comparing a coherent signal model modes to incoherent signal models. For the F-statistic and the 5n-vector method, the columns show the p-value (for the 5n-vector method, considering a signal at just the $l = 2\,, m = 1$ mode for the dual-harmonic emission model).
}
\tablerefs{The following is a list of references for pulsar ephemeris data used in this analysis: Nancay: $\alpha$, NICER: $\beta$, JBO: $\gamma$, IAR: $\delta$, Hobart: $\epsilon$, FAST: $\zeta$, CHIME: $\eta$, Chandra: $\theta$, Fermi-LAT: $\kappa$.}
\tablerefs{The following is a list of references for pulsar distances and intrinsic period derivatives, and they should be consulted for information on the associated uncertainties on these quantities: (a) \cite{2023MNRAS.519.4982D}, (b) \cite{2004AA...415..531S}, (c) \cite{2017ApJ...835...29Y}, (d) \cite{1993AA...274..427R}, (e) \cite{2016MNRAS.455.1751R}, (f) \cite{1968AJ.....73..535T}, (g) \cite{2012ApSS.341...43W}, (h) \cite{2016MNRAS.455.3806B}, (i) \cite{2012ApJ...755...39V}, (j) \cite{2021ApJ...922..253M}, (k) \cite{2021ApJ...921L..19D}, (l) \cite{2021MNRAS.505.5957B}, (m) \cite{2014MNRAS.443.2183F}, (n) \cite{1988MNRAS.231..735G}, (o) \cite{2019MNRAS.489.5494T}, (p) \cite{2009MNRAS.400..168L}, (q) \cite{2018AA...612A...2H}, (r) \cite{2024MNRAS.528.6761L}, (s) \cite{2015ApJ...802...17K}, (t) \cite{2011ApJ...739...39A}, (u) \cite{2016AA...587A.109G}, (v) \cite{2021AA...654A..16G}, (w) \cite{2001ApJ...547..323H}.}
\end{deluxetable}
\end{longrotatetable}

\begin{table*}[ht]
    \centering
    \caption{Upper limits on the strain amplitude set with the narrowband search for each of the considered targets. As a reference, we list also the number of templates, N$_{\mathrm{trials}}$, in the $f-\Dot{f}$ plane. We report as well the lowest p-value found in the analysis with the corresponding threshold set after correcting a FAP of $10^{-2}$ for the trial factor. The nomenclature "pg" identifies the post-glitch analysis.}
    \label{tab:narrowband_results}
    \renewcommand*{\arraystretch}{1.2}
    \begin{tabular}{cccccc}
    \hline\hline
        Pulsar Name & N$_{\mathrm{trials}}$ & $h_0^{95\%}$ & $h_0^{95\%}/h_0^{sd}$ & Lowest p-value & Threshold p-value\\
        (J2000) & $\times 10^6$ & $\times 10^{-26}$ & & $\times 10^{-5}$ & $\times 10^{-11}$\\
         \hline
        J0205+6449 & 90 & 9.60 & 0.22 & 1.54 & 1.10 \\
        J0534+2200 & 1383 & 4.16 & 0.03 & 0.77 & 0.07 \\
        J0537-6910 & 298 & 3.52 & 1.21 & 0.13 & 0.33 \\
        J0537-6910 pg & 119 & 3.75 & 1.29 & $4\cdot10^{-4}$ & 0.84 \\
        J0540-6919 & 11 & 14.23 & 2.45 & 0.44 & 8.95 \\
        J0540-6919 pg & 248 & 7.38 & 1.27 & 0.02 & 0.40 \\
        J0835-4510 & 23 & 21.53 & 0.06 & 8.50 & 4.18 \\
        J1811-1925 & 23 & 9.71 & 0.72 & 0.45 & 4.29 \\
        J1813-1246 & 24 & 5.81 & 0.31 & 4.07 & 4.03 \\
        J1813-1749 & 182 & 5.17 & 0.23 & 0.04 & 0.55 \\
        J1838-0655 & 21 & 12.98 & 1.27 & 2.93 & 4.68 \\
        J1913+1011 & 15 & 4.14 & 0.77 & 1.40 & 6.45 \\
        J1935+2025 & 16 & 17.84 & 1.17 & 0.63 & 5.94 \\
        J1952+3252 & 18 & 4.24 & 0.41 & 1.06 & 5.55 \\
        J2016+3711 & 73 & 6.31 & 0.40 & 1.37 & 1.37 \\
        J2021+3651 & 12 & 37.11 & 0.86 & 0.25 & 7.73 \\
        J2022+3842 & 99 & 5.80 & 0.54 & 0.25 & 1.01 \\
        J2229+6114 & 77 & 6.19 & 0.19 & 0.62 & 1.28 \\
         \hline
    \end{tabular}
\end{table*}

%\startlongtable
%\begin{longrotatetable}
\begin{deluxetable*}{lrlccc}
\tablecaption{Limits on GW amplitude from dipole radiation in Brans-Dicke theory.
\label{tab:BDresults}}
\tabletypesize{\scriptsize}
\setlength{\tabcolsep}{3.5pt}
\tablehead{
\colhead{Pulsar Name} &
\colhead{$f_{\text{rot}}$} &
\colhead{$\dot{f}_{\text{rot}}$} &
\colhead{Distance} &
\colhead{$h_{0d}^{95\%}$} &
\colhead{FAP}\\
\colhead{(J2000)} &
\colhead{(Hz)} &
\colhead{(Hz s$^{-1})$} &
\colhead{(kpc)} &
\colhead{~} &
\colhead{~}}
\startdata
J0030+0451\tablenotemark{$\alpha$} & 205.53 & \scinum{-4.23}{-16} & 0.33\tablenotemark{a} & \scinum{8.77}{-27} & 0.98  \\
J0058\textminus7218\tablenotemark{$\beta$} & 45.94 & \scinum{-6.1}{-11} & 59.70\tablenotemark{b} & \scinum{3.02}{-26} & 0.73 \\
J0117+5914\tablenotemark{$\gamma$} & 9.86 & \scinum{-5.69}{-13} & 1.77\tablenotemark{c} & \scinum{2.03}{-23} & 1 \\
J0205+6449\tablenotemark{$\gamma$} & 15.22 & \scinum{-4.49}{-11} & 3.20\tablenotemark{d} & \scinum{3.78(2.57)}{-25} & 0.99(0.79) \\
J0437\textminus4715\tablenotemark{$\delta$} & 173.69 & \scinum{-4.15}{-16} & 0.16\tablenotemark{e} & \scinum{8.13}{-27} & 0.98  \\
J0534+2200\tablenotemark{$\gamma$} & 29.95 & \scinum{-3.78}{-10} & 2.00\tablenotemark{f} & \scinum{1.37(0.86)}{-25} & 0.46(0.99) \\
J0537\textminus6910\tablenotemark{$\beta$} & 62.03 & \scinum{-1.99}{-10} & 49.70\tablenotemark{g} & \scinum{1.84(1.11)}{-26} & 0.98(0.60) \\
J0540\textminus6919\tablenotemark{$\beta$} & 19.77 & \scinum{-1.87}{-10} & 49.70\tablenotemark{g} & \scinum{2.41(1.44)}{-25} & 0.92(0.36) \\
J0614\textminus3329\tablenotemark{$\alpha$} & 317.59 & \scinum{-1.76}{-15} & 0.63\tablenotemark{h} & \scinum{1.78}{-26} & 0.56  \\
J0737\textminus3039A\tablenotemark{$\alpha$} & 44.05 & \scinum{-3.41}{-15} & 1.10\tablenotemark{i} & \scinum{4.08}{-26} & 0.83 \\
J0835\textminus4510\tablenotemark{$\epsilon$} & 11.19 & \scinum{-1.57}{-11} & 0.28\tablenotemark{i} & \scinum{4.12(2.73)}{-24} & 0.80(0.33) \\
J1231\textminus1411\tablenotemark{$\alpha$} & 271.45 & \scinum{-5.92}{-16} & 0.42\tablenotemark{c} & \scinum{3.21}{-26} & 1 \\
J1412+7922\tablenotemark{$\beta$} & 17.18 & \scinum{-9.72}{-13} & 3.30\tablenotemark{j} & \scinum{5.56}{-26} & 0.93  \\
J1537+1155\tablenotemark{$\alpha$} & 26.38 & \scinum{-1.65}{-15} & 0.93\tablenotemark{k} & \scinum{5.93}{-26} & 0.97 \\
J1623\textminus2631\tablenotemark{$\gamma$} & 90.29 & \scinum{-5.26}{-15} & 1.85\tablenotemark{l} & \scinum{3.10}{-25} & 0.77  \\
J1719\textminus1438\tablenotemark{$\alpha$} & 172.71 & \scinum{-2.22}{-16} & 0.34\tablenotemark{c} & \scinum{5.44}{-27} & 1  \\
J1744\textminus1134\tablenotemark{$\alpha$} & 245.43 & \scinum{-4.34}{-16} & 0.40\tablenotemark{e} & \scinum{1.60}{-26} & 0.92   \\
J1745\textminus0952\tablenotemark{$\alpha$} & 51.61 & \scinum{-2.3}{-16} & 0.23\tablenotemark{c} & \scinum{4.30}{-26} & 0.81  \\
J1756\textminus2251\tablenotemark{$\gamma$} & 35.14 & \scinum{-1.26}{-15} & 0.73\tablenotemark{m} & \scinum{5.29}{-26} & 0.64   \\
J1809\textminus1917\tablenotemark{$\gamma$} & 12.08 & \scinum{-3.73}{-12} & 3.27\tablenotemark{c} & \scinum{3.25}{-24} & 0.90   \\
J1811\textminus1925\tablenotemark{$\beta$} & 15.46 & \scinum{-1.05}{-11} & 5.00\tablenotemark{n} & \scinum{6.20}{-25} & 0.98  \\
J1813\textminus1246\tablenotemark{$\beta$} & 20.80 & \scinum{-7.6}{-12} & 2.63\tablenotemark{o} & \scinum{3.60}{-25} & 0.44 \\
J1813\textminus1749\tablenotemark{$\beta$} & 22.35 & \scinum{-6.34}{-11} & 6.15\tablenotemark{c} & \scinum{1.09}{-25} & 0.97 \\
J1823\textminus3021A\tablenotemark{$\gamma$} & 183.82 & \scinum{-1.14}{-13} & 8.02\tablenotemark{l} & \scinum{1.84}{-26} &  0.58  \\
J1824\textminus2452A\tablenotemark{$\alpha$} & 327.41 & \scinum{-1.73}{-13} & 5.37\tablenotemark{l} & \scinum{2.08}{-26} & 0.49  \\
J1826\textminus1334\tablenotemark{$\gamma$} & 9.85 & \scinum{-7.31}{-12} & 3.61\tablenotemark{c} & \scinum{3.69}{-25} &  0.98 \\
J1828\textminus1101\tablenotemark{$\gamma$} & 13.88 & \scinum{-2.85}{-12} & 4.77\tablenotemark{c} & \scinum{3.68}{-24} & 1 \\
J1831\textminus0952\tablenotemark{$\gamma$} & 14.87 & \scinum{-1.84}{-12} & 3.68\tablenotemark{c} & \scinum{1.13}{-24} & 0.87  \\
J1833\textminus0827\tablenotemark{$\gamma$} & 11.72 & \scinum{-1.26}{-12} & 4.50\tablenotemark{i} & \scinum{3.79}{-24} & 0.82  \\
J1837\textminus0604\tablenotemark{$\gamma$} & 10.38 & \scinum{-4.84}{-12} & 4.78\tablenotemark{c} & \scinum{9.77}{-24} & 0.91  \\
J1838\textminus0655\tablenotemark{$\beta$} & 14.18 & \scinum{-9.9}{-12} & 6.60\tablenotemark{p} & \scinum{2.91}{-24} &  0.51\\
J1849\textminus0001\tablenotemark{$\beta$} & 25.96 & \scinum{-9.54}{-12} & 7.00\tablenotemark{q} & \scinum{1.70}{-25} & 0.34   \\
J1856+0245\tablenotemark{$\gamma$} & 12.36 & \scinum{-9.49}{-12} & 6.32\tablenotemark{c} & \scinum{7.32}{-25} & 1 \\
J1913+1011\tablenotemark{$\gamma$} & 27.85 & \scinum{-2.62}{-12} & 4.61\tablenotemark{c} & \scinum{8.43}{-26} &  0.88 \\
J1925+1720\tablenotemark{$\gamma$} & 13.22 & \scinum{-1.83}{-12} & 5.05\tablenotemark{c} & \scinum{4.00}{-24} &  0.70 \\
J1935+2025\tablenotemark{$\gamma$} & 12.48 & \scinum{-9.47}{-12} & 4.60\tablenotemark{c} & \scinum{1.57}{-24} &  0.95\\
J1952+3252\tablenotemark{$\gamma$} & 25.30 & \scinum{-3.74}{-12} & 3.00\tablenotemark{i} & \scinum{1.01}{-25} &  0.78 \\
J2016+3711\tablenotemark{$\zeta$} & 19.68 & \scinum{-2.81}{-11} & 6.10\tablenotemark{r} & \scinum{1.91}{-25} &  0.90 \\
J2021+3651\tablenotemark{$\gamma$} & 9.64 & \scinum{-8.89}{-12} & 1.80\tablenotemark{s} & \scinum{1.32(0.61)}{-22} & 0.39(0.08)\\
J2022+3842\tablenotemark{$\beta$} & 20.59 & \scinum{-3.65}{-11} & 10.00\tablenotemark{t} & \scinum{7.44}{-24} &   0.67 \\
J2043+2740\tablenotemark{$\gamma$} & 10.40 & \scinum{-1.37}{-13} & 1.48\tablenotemark{c} & \scinum{3.93}{-24} &  0.97 \\
J2214+3000\tablenotemark{$\eta$} & 320.59 & \scinum{-1.31}{-15} & 0.60\tablenotemark{u} & \scinum{5.68}{-27} &  1 \\
J2222\textminus0137\tablenotemark{$\alpha$} & 30.47 & \scinum{-4.99}{-16} & 0.27\tablenotemark{v} & \scinum{5.21}{-26} & 0.96  \\
J2229+6114\tablenotemark{$\gamma$} & 19.36 & \scinum{-2.9}{-11} & 3.00\tablenotemark{w} & \scinum{2.31(1.61)}{-25} &  0.86(0.25) \\
\enddata
\tablecomments{For references and other notes see Table~\ref{tab:results}. Values in parentheses are those produced using the restricted orientation priors described in Section~\ref{sec:restricted}. The last column shows the false-alarm probability (FAP) for a signal, assuming that the $2\mathcal{D}$ value has a $\chi^2$ distribution with 2 degrees-of-freedom.}
\end{deluxetable*}
%\end{longrotatetable}

%\clearpage
 
\bibliographystyle{aasjournal}
\bibliography{bibliography}
\end{document}